\documentclass[a4paper,11pt]{article}
\pdfoutput=1 

\usepackage{jheppub} 
\usepackage{amsmath,amssymb}
\usepackage{datetime,mathdots}
\usepackage{verbatim}
\usepackage[utf8]{inputenc}

\def\be{\begin{equation}}
\def\ee{\end{equation}}
\def\bea{\begin{eqnarray}}
\def\eea{\end{eqnarray}}
\def\bal{\begin{align}}
\def\eal{\end{align}}

\newcommand\bR{\mathbb{R}}

\newcommand\bS{\mathbb{S}}
\newcommand\bZ{\mathbb{Z}}
\newcommand\bC{\mathbb{C}}

\newcommand\cN{\mathcal{N}}
\newcommand\cL{\mathcal{L}}

\newcommand\cA{\mathcal{A}}

\newcommand\cC{\mathcal{C}}

\newcommand\cH{\mathcal{H}}

\newcommand\cM{\mathcal{M}}

\newcommand\cT{\mathcal{T}}

\newcommand\cW{\mathcal{W}}
\newcommand\cX{\mathcal{X}}

\newcommand\ex{\mathrm{e}}
\newcommand\ii{\mathrm{i}}

\newcommand\qqq{\qquad\qquad}
\newcommand{\nn}{\nonumber \\ {} }

\DeclareMathOperator{\im}{Im}
\DeclareMathOperator{\re}{Re}
\newcommand{\Tr}{\mathrm{Tr}}
\newcommand{\Hol}{\mathrm{Hol}}
\newcommand{\tr}{\mathrm{tr}}
\newcommand{\ts}[1]{\textsuperscript{\tiny #1}}
\renewcommand{\th}{\ts{th}\ }

\newcommand{\nab}{\nabla^{\text{ab}}}

\title{Quantum Holonomies from Spectral Networks and Framed BPS States}

\author[a]{Maxime Gabella}

\affiliation[a]{Institute for Advanced Study, Einstein Drive, Princeton, New Jersey 08540, USA}

\emailAdd{gabella@ias.edu}

\abstract{
We propose a method for determining the spins of BPS states supported on line defects in 4d $\mathcal{N}=2$ theories of class~S. Via the 2d-4d correspondence, this translates to the construction of quantum holonomies on a punctured Riemann surface $\mathcal{C}$. Our approach combines the technology of spectral networks, which decomposes flat $GL(K,\bC)$-connections on $\mathcal{C}$ in terms of flat abelian connections on a $K$-fold cover of $\mathcal{C}$, and the skein algebra in the 3-manifold $\mathcal{C}\times [0,1]$, which expresses the representation theory of the quantum group $U_q(gl_K)$. With any path on $\mathcal{C}$, the quantum holonomy associates a positive Laurent polynomial in the quantized Fock-Goncharov coordinates of higher Teichm\"uller space. This confirms various positivity conjectures in physics and mathematics.
}

\begin{document} 

\maketitle
\flushbottom

\newpage

%%%%%%%%%%%%%%
\section{Introduction}

The maximally supersymmetric conformal field theories in the maximum number of dimensions are the six-dimensional $(2,0)$ superconformal theories, which are labeled by a simply-laced Lie algebra $\mathfrak{g}$. The $(2,0)$ theory with $\mathfrak{g} = A_{K-1}$ can be realized in M-theory as the low-energy limit of the worldvolume theory of $K$ coincident M5-branes.
Although little is known about these theories, their existence leads to a geometric description of many supersymmetric field theories in lower dimensions.
As a prime example, the twisted compactification of the six-dimensional $(2,0)$ theories on a Riemann surface $\cC$ produces a rich class of $\cN=2$ supersymmetric theories in four dimensions (called \emph{class S}), whose properties are encoded in the geometry of $\cC$~\cite{Gaiotto:2009hg, Gaiotto:2009we, Alday:2009aq}.

The 6d $(2,0)$ theory contains dynamical strings, which arise from the two-dimensional boundaries of open M2-branes stretched between M5-branes.
In the limit where its tension tends to infinity, a dynamical string becomes a nonabelian surface defect $\bS$, which can preserve some supersymmetry.
After compactification on $\cC$, a surface defect extending along the time direction in 4d spacetime and wrapping a closed path $\wp$ on $\cC$ defines a line defect $L_\wp$ in the 4d $\cN=2$ theory, such as a Wilson or 't Hooft operator~\cite{Drukker:2009tz, Alday:2009fs, Drukker:2009id}.
The expectation value of a supersymmetric line defect $L_\wp$ is conjectured to admit a decomposition of the form~\cite{Gaiotto:2010be}
\bea \label{LUVdecomp}
\langle L_\wp \rangle = \sum_\gamma \underline{\overline{\Omega}}(\wp,\gamma) \cX_\gamma~,
\eea
where $\cX_\gamma$ can be thought of as the expectation value of a line defect with electromagnetic charge $\gamma$ in the low-energy IR theory, in which the $SU(K)$ gauge symmetry is broken to the abelian subgroup $U(1)^{K-1}$.
The coefficients $\underline{\overline{\Omega}}(\wp,\gamma)$ are integers which count the BPS states supported on $L_\wp$, called \emph{framed BPS states}.
It is possible to account for the spins of framed BPS states by refining this BPS index into the \emph{framed protected spin character} $\underline{\overline{\Omega}}(\wp,\gamma;q)$, which is a function of a variable~$q$.
It is defined as a trace over the Hilbert space $\cH_{L,\gamma}$ of framed BPS states with charge $\gamma$:
\bea \label{framedPSCdef}
\underline{\overline{\Omega}}(\wp,\gamma;q)  =  \Tr_{\cH_{L,\gamma}} q^{2J_3} (-q)^{2I_3}~,
\eea
where $J_3$ and $I_3$ are Cartan generators of the spatial rotations $so(3)$ and of the $su(2)_R$ symmetry (it reduces to the framed BPS index $\underline{\overline{\Omega}}(\wp,\gamma)$ for $q=-1$).
According to the ``no exotics conjecture,'' all BPS states are in fact singlets of $su(2)_R$ with $I_3=0$.
The generating function of framed protected spin characters is 
\bea\label{GenFctPSC}
F(L_\wp) = \sum_\gamma\underline{\overline{\Omega}}(\wp,\gamma;q) \hat \cX_\gamma~,
\eea
where the noncommutative variables $\hat \cX_\gamma$ satisfy the relation
\bea \label{XgNoncom}
\hat \cX_{\gamma_1} \hat \cX_{\gamma_2} = q^{\langle \gamma_1, \gamma_2 \rangle} \hat \cX_{\gamma_1\gamma_2}~,
\eea
with $\langle \cdot, \cdot \rangle$ denoting the Dirac-Schwinger-Zwanziger antisymmetric product of charges.

A method for computing the framed BPS indices $\underline{\overline{\Omega}}(\wp,\gamma)$ for 4d $\cN=2$ theories of class S was provided in~\cite{Gaiotto:2010be}.
The moduli space $\cM$ of vacua of these theories on $\bR^3 \times S^1$ is isomorphic to the hyperk\"ahler moduli space of solutions to Hitchin's equations on $\cC$ with singularities at the punctures~\cite{Gaiotto:2008cd, Gaiotto:2009hg}.
These equations concern a connection $A$ and a 1-form $\varphi$, and amount to the flatness of the complex connections 
\bea
\cA(\zeta) = R \frac{\varphi}{\zeta} + A + R \zeta \bar \varphi
\eea
for all values of $\zeta \in \bC^\times$, where $R$ is the radius of the compactification circle $S^1$.
In fact, in one of its complex structure, $\cM$ can be identified with the moduli space of complex flat connections $\cA(\zeta)$ with certain monodromies around the punctures.
The functions $\cX_\gamma$ in~\eqref{LUVdecomp} are Darboux coordinates on $\cM$, closely related to the Fock-Goncharov coordinates of higher Teichm\"uller theory,  associated with ideal triangulations of $\cC$~\cite{FG}.

The expectation value of the line defect $L_\wp$ can then be expressed  as the holonomy of a flat connection $\cA$ along the path $\wp$~\cite{Gaiotto:2010be}:
\bea
\langle L_\wp\rangle = \Tr\, \text{Hol}_\wp \cA~.
\eea
There is a well-known algorithm to expand these holonomies in terms of Fock-Goncharov coordinates~\cite{FG}. 
The decomposition~\eqref{LUVdecomp} of UV line defects in terms of IR line defects thus corresponds to a map from the space of paths $\wp$ on $\cC$ to the space of Laurent polynomials in the coordinates $\cX_\gamma$ on the moduli space of flat connections.
The framed BPS indices $\underline{\overline{\Omega}}(\wp,\gamma)$ are the coefficients in these polynomials.

In this paper, we propose a method for computing the framed protected spin characters $\underline{\overline{\Omega}}(\wp,\gamma;q) $ by constructing quantum holonomies of flat $PGL(K,\bC)$-connections,  expressed as Laurent polynomials in the quantized Fock-Goncharov coordinates.
We achieve this mainly by combining two recent developments: the ``spectral networks with spin'' of Galakhov,  Longhi, and Moore~\cite{Galakhov:2014xba}, and the $SL(2)$ ``quantum trace'' of Bonahon and Wong~\cite{2010arXiv1003.5250B}.

A quantum parallel transport along open paths on $\cC$ was defined in~\cite{Galakhov:2014xba} using the technology of \emph{spectral networks}~\cite{Gaiotto:2012rg,Gaiotto:2012db}. In their simplest incarnation, spectral networks are collections of trajectories (walls) starting from the branch points of a $K$-fold branched covering $\Sigma\to \cC$ (the Seiberg-Witten curve or Hitchin spectral curve) and ending at the punctures (section~\ref{secSpecNet}). They lead to a \emph{nonabelianization map}, by which flat $GL(K,\bC)$-connections on $\cC$ can be described in terms of flat abelian connections on $\Sigma$ (section~\ref{secNonab}).
The quantum parallel transport was constructed in~\cite{Galakhov:2014xba} by identifying the spin $2J_3$ of a framed BPS state of charge $\gamma$ with the writhe, that is a signed sum over self-intersections, of the corresponding path $\gamma$ on $\Sigma$ (section~\ref{secQPT}). This gives an expression like~\eqref{GenFctPSC}, where $\langle \cdot, \cdot \rangle$ now denotes the intersection pairing on $H_1(\Sigma,\bZ)$.
However, this quantum parallel transport cannot be used as such to compute quantum holonomies for \emph{closed} paths $\wp$ because the result would depend on the choice of basepoint.
Indeed, moving the basepoint across a self-intersection of a path $\gamma$ in $\Sigma$ changes its handedness, and hence its contribution to the spin (it exchanges $q$ and $q^{-1}$).

The issue of the dependence on the choice of basepoint can be solved by thinking of $\wp$ as a closed path (a knot) in the 3-manifold
\bea \label{M3CI}
M_3 = \cC \times [0,1]~.
\eea
The elevation along the interval $[0,1]$ then corresponds to a quantum ordering.
The algebra of line defects in the $\cN=2$ theory on $ \bR^3\times S^1$ is indeed quantized by twisting to the fibered product $\bR \times \bR^2 \times_q S^1$ such that the coordinates on $\bR^2$ are rotated by $q$ after going around~$S^1$.
BPS line defects on $S^1$ then have a definite ordering along the axis~$\bR$, which thus plays the same role as the interval $[0,1]$ in~\eqref{M3CI}.
Moreover, motivated by the relation with Chern-Simons theory~\cite{WittenJones, Witten:1989wf}, Turaev showed that the Poisson algebra of paths on a surface $\cC$ can be quantized by the skein algebra of knots in $M_3$~\cite{Turaev:1991}.
This 3-dimensional approach was used successfully by Bonahon and Wong to define a \emph{quantum trace} for $SL(2,\bC)$~\cite{2010arXiv1003.5250B} (see also~\cite{2015arXiv151106054L, Allegretti:2015nxa}), that is a homomorphism from the skein algebra to the quantum Teichm\"uller space~\cite{Chekhov:1999tn}. The quantum trace maps every isotopy class of knot in $M_3$ to a Laurent polynomial in the quantized shear coordinates associated with an ideal triangulation of $\cC$. 
The isotopy invariance of the quantum trace is ensured by carefully controlling the  elevation of segments of the knot, which is achieved by inserting certain transformations over the edges of the triangulation.
These transformations are related to the R-matrix and to the cup/cap matrices of the quantum group $U_q(sl_2)$, such as those used by Reshetikhin and Turaev to construct isotopy invariants of knots~\cite{RT1}.
The appearance of the quantum R-matrix is natural from the perspective of Chern-Simons theory, where it corresponds to a crossing of two Wilson lines~\cite{Witten:1989wf}. 

Inspired by these developments, we construct a quantum holonomy for flat $PGL(K,\bC)$-connections which associates with every path $\wp$ on $\cC$ a Laurent polynomial in the quantized Fock-Goncharov coordinates $\hat x_\alpha$:
\bea \label{TrHolqwp}
\Tr\, \Hol^q_\wp = \sum_\gamma\underline{\overline{\Omega}}(\wp,\gamma;q) \hat \cX_\gamma~,
\eea 
where the coefficients $\underline{\overline{\Omega}}(\wp,\gamma;q)$ are the framed protected spin characters, and $\hat \cX_\gamma$ is of the form
\bea 
\hat \cX_\gamma =   q^{- \sum_{\alpha <\beta}a_\alpha \varepsilon_{\alpha\beta}  a _\beta } \hat x_1^{a_1} \cdots \hat x_n^{a_n}~
\eea
for some integers $a_\alpha$.
The sum in~\eqref{TrHolqwp} is over paths $\gamma$ on the $K$-fold cover $\Sigma$ that are lifted from $\wp$ on $\cC$, with certain \emph{detours} along the walls of the spectral network.
The strategy is to first compute the quantum parallel transport along segments of $\wp$ on each triangle of the ideal triangulation, and then to glue them together with $U_q(gl_K)$-matrices to control their relative elevation along the interval $[0,1]$ (section~\ref{secQHol}).
This guarantees that the quantum holonomy is invariant under isotopy in $M_3$, and so in particular under changes in the choice of basepoint (thought of as the lowest point of $\wp$ along $[0,1]$).
By construction, the coefficients $\underline{\overline{\Omega}}(\wp,\gamma;q)$ are Laurent polynomials in~$q$. In section~\ref{secProp}, we show that they are in fact \emph{positive} Laurent polynomials, and are invariant under inversion of $q$.
This is in agreement with the positivity conjectures by Gaiotto, Moore, and Neitzke~\cite{Gaiotto:2010be}, and with the related conjectures by Fock and Goncharov~\cite{FG, 2003math.....11245F}.

The properties of the quantum holonomy are clearest for a simple path $\wp$ (without self-intersection), or more generally for a lamination, that is a collection of non-intersecting simple paths. 
When $\wp$ has self-intersections, the invariance under inversion of $q$ is spoiled.
However, skein relations such as those described in section~\ref{secSkeinRel} resolve intersections into non-intersecting paths and pairs of trivalent junctions. 
The quantum holonomy for the resulting \emph{network}\footnote{
Such networks should not be confused with spectral networks!}
with junctions enjoys the same nice properties as for a simple path. 

Our results can be formalized in the following theorem:

\paragraph{Theorem 1:}\textit{
There exists a quantum holonomy map from the space $\cA_K$ of simple paths and networks on a Riemann surface $\cC$ to the quantized algebra $\cX^q_K$ of functions on the moduli space of flat $PGL(K,\bC)$-connections on $\cC$:
\bea
\Tr\, \Hol^q :   \quad \cA_K   \to \cX^q_K~.
\eea
It associates with any simple path or network $\wp$ on $\cC$ a Laurent polynomial 
\bea
\Tr\, \Hol^q_\wp = \sum_\gamma\underline{\overline{\Omega}}(\wp,\gamma;q) \hat \cX_\gamma~
\eea 
with the following properties:
\begin{enumerate}
\item 
It agrees with the classical trace of the holonomy of a flat $PGL(K,\bC)$-connection along $\wp$ when $q=1$:
\bea
\Tr\, \Hol^1_\wp = \Tr\, \Hol_\wp~.
\eea
\item 
Its highest term has a unit coefficient:
\bea
\underline{\overline{\Omega}}(\wp,\gamma^\mathrm{highest} ;q)  = 1~,
\eea
\item
Its coefficients are positive Laurent polynomials in $q$:
\bea
\underline{\overline{\Omega}}(\wp,\gamma;q) \; \in\;  \bZ_{>0}[q,q^{-1}]
\eea
\item
It is invariant under inversion of $q$:
\bea
\underline{\overline{\Omega}}(\wp,\gamma;q) = \underline{\overline{\Omega}}(\wp,\gamma;q^{-1})~.
\eea
\end{enumerate}
}

\

\noindent
Property~3 confirms the ``weak positivity conjecture'' in~\cite{Gaiotto:2010be}, and furthermore Property~4 brings support to the ``strong positivity conjecture.''
To make precise contact with the conjectures of Fock and Goncharov (Conjecture~12.4 in~\cite{FG} and Conjecture~4.8 in~\cite{2003math.....11245F}) would require a better understanding of the relation between the space of tropical points of the so-called $\cA$-space and the space of networks with junctions (see for example the ``higher laminations'' of~\cite{2012arXiv1209.0812L, Xie:2013lca} and the ``charge/network dictionary'' in~\cite{Tachikawa:2015iba, Watanabe:2016bwr}).
It should also be possible to relate exactly to the $SL(2,\bC)$ quantum trace of Bonahon and Wong~\cite{2010arXiv1003.5250B} by introducing some normalizations, and to see that their application of Weyl quantum ordering matches the pattern of intersections of detoured paths.

We illustrate our construction of quantum holonomies in section~\ref{secExamples} by computing some framed protected spin characters for 4d $\cN=2$ theories associated with the punctured torus and with the three-punctured sphere.
They agree with the ones deduced in~\cite{Coman:2015lna} from the fact that the noncommutative algebra of line operators coincides with the skein algebra.
It would also be interesting to compare our method for computing framed protected spin characters to the method based on representations of framed BPS quivers~\cite{Chuang:2013wt, Cirafici:2013bha, Cordova:2013bza}.

Note that the quantized algebra of functions on the moduli space of flat $PGL(K,\bC)$-connections was shown in~\cite{Bullimore:2013xsa, Coman:2015lna} to match the algebra of Verlinde loop and network operators in $SU(K)$ Toda field theory.
This follows essentially from the observation that the braiding matrix, from which the Verlinde operators are built, is related to the quantum R-matrix of $U_q(sl_K)$.
Nonabelianization certainly has an interesting interpretation in Toda theory too.

Our three-dimensional construction of quantum holonomies should find a natural environment in the 3d-3d correspondence, which relates 3d $\cN=2$ supersymmetric field theories to $SL(K,\bC)$ Chern-Simons theory on a 3-manifold~\cite{Dimofte:2011ju, Cecotti:2011iy, Dimofte:2013iv} (see~\cite{Dimofte:2014ija} for a review).
We also anticipate exciting connections with the upcoming work on 3d spectral networks by Freed and Neitzke~\cite{FreedNeitzke}.

%%%%%%%%%%%%%%
%%%%%%%%%%%%%%
\section{Spectral networks} \label{secSpecNet}

A spectral network is a collection of paths on a punctured Riemann surface $\cC$ obeying certain local conditions.
Spectral networks were introduced in~\cite{Gaiotto:2012rg,Gaiotto:2012db} and shown to play a fundamental role in understanding the spectrum of BPS states and wall-crossing in $\cN=2$ supersymmetric field theories of class S.
They provide a relation between flat $GL(K,\bC)$-connections on $\cC$ and abelian flat connections on a $K$-fold branched cover $\Sigma$ of $\cC$.
We first present their abstract definition and then review the physical motivations behind it.
We also describe the special family of spectral networks on which we will focus in this paper, because of their relation with ideal triangulations of $\cC$.

%%%%%%%%%%%%%%
\subsection{Abstract definition}

Let $\cC$ be an oriented Riemann surface with a non-empty set of marked points and a (possibly empty) boundary. Marked points in the interior of $\cC$ are referred to as \emph{punctures}, and each boundary component must contain at least one marked point. 
Let $\pi: \Sigma\to \cC$ be a $K$-fold branched covering of $\cC$, which is unramified over the boundary and the punctures. 
It is convenient to choose a set of branch cuts on $\cC$, on the complement of which the covering can be trivialized and the sheets of $\Sigma$ labeled by integers $i = 1, \ldots , K$.
The branch points are assumed to be simple, so that a monodromy around a branch point of type $(ij)$ exchanges sheet $i$ and sheet $j$.

A \emph{spectral network $\cW$} subordinate to the covering $\Sigma\to \cC$ is a collection of oriented paths on $\cC$, called \emph{walls}, labeled by ordered pairs $ij$. 
Exactly three walls (of types $ij$ or $ji$) begin at each branch point of type $(ij)$, and each wall ends at a puncture or at a marked point on the boundary. In degenerate cases, a pair of oppositely-oriented walls can be stretched between two branch points.
An $ij$-wall can also begin or end at the intersection of an $ik$-wall and a $kj$-wall.

%%%%%%%%%%%%%%
\subsection{Spectral networks in $\cN=2$ theories} \label{secSpecNetN2Th}

Spectral networks arise naturally in $\cN=2$ supersymmetric field theories in four dimensions. We focus on theories of class S associated with the Lie algebra $A_{K-1}$, which describe a system of $K$ coincident M5-branes wrapping a punctured Riemann surface~$\cC$. At a generic point of the Coulomb branch of the 4d $\cN=2$ theory, the gauge symmetry group is abelian in the IR, and the M5-branes separate by wrapping a $K$-fold branched cover $\Sigma$ of $\cC$.
This is the Seiberg-Witten curve, given by the spectral curve of an $A_{K-1}$ Hitchin system:
\bea \label{SigmaCurve}
\Sigma = \{ \lambda \in  T^* \cC: \; \lambda^K + \sum_{r=2}^K \phi_r \lambda^{K-r} = 0 \}  ~ ,
\eea
where $\phi_r$ are meromorphic $r$-differentials on $\cC$ with prescribed singularities at the punctures.
The restriction of the Liouville one-form to $\Sigma$ gives a natural holomorphic one-form, which we also denote by $\lambda$.
Choosing a trivialization of $\Sigma \to \cC$ corresponds to labeling the solutions of~\eqref{SigmaCurve} as $\lambda_i$, with $i=1, \ldots, K$, so that the graph of $\lambda_i$ in $T^*\cC$ is sheet $i$ of $\Sigma$.
A branch point of type $(ij)$ is a point $z\in \cC$ where sheets $i$ and $j$ collide, $\lambda_i(z)= \lambda_j(z)$.

An $ij$-trajectory with phase $\vartheta$ is a path $p$ on $\cC$ satisfying the differential condition
\bea \label{ijTraj}
\im \left[ \ex^{-\ii\vartheta} \langle \lambda_{ij}, v \rangle \right] = 0 ~,
\eea
where $\lambda_{ij}=\lambda_{i} -\lambda_{j}$, and $v$ is a vector field along $p$. 
It has a natural orientation, with the positive direction corresponding to $\re \left[ \ex^{-\ii\vartheta} \langle \lambda_{ij}, v \rangle \right] >0$.
The walls of a spectral network are certain $ij$-trajectories.
Before giving a more precise definition, we review the relevance of the condition~\eqref{ijTraj} for various BPS states. 

BPS states in the 4d theory arise from strings in the 6d theory that extend along paths $p$ on $\cC$, and hence look like point-particles.
These paths $p$ are labeled by pairs $ij$ of sheets and lift to closed paths $\gamma$ on $\Sigma$, which can be thought of as boundaries of M2-branes stretched between M5-branes. This implies that the paths $p$ may be closed paths, or have endpoints at branch points and at junctions with other strings. For example, an open path $p$ with label $ij$ between two branch points on $\cC$ lifts to a closed path $\gamma=p^{(i)}(p^{(j)})^{-1}$  on $\Sigma$, where $p^{(i)}$ denotes the lift of $p$ to sheet $i$, and the minus sign means reverse orientation.
The homology class of $\gamma$ in $H_1(\Sigma; \bZ)$ is the charge of the corresponding state, while its central charge and mass are given by
\bea \label{ZMlambda}
Z =  \frac 1\pi \int_p \lambda_{ij}  ~, \qqq  M = \frac 1\pi \int_p |\lambda_{ij} |  ~.
\eea
The BPS bound $|Z| \le M$ is saturated when $\lambda_{ij} $ has the same phase $\vartheta$ everywhere along $p$, in which case $Z = \ex^{\ii \vartheta} M$.
This is true if and only if the condition~\eqref{ijTraj} is satisfied.
BPS states therefore correspond to webs of $ij$-trajectories. The value of of the parameter $\zeta = \ex^{\ii \vartheta}$ determines which supercharges are preserved.

\begin{figure}[tb]
\centering
\includegraphics[width=\textwidth]{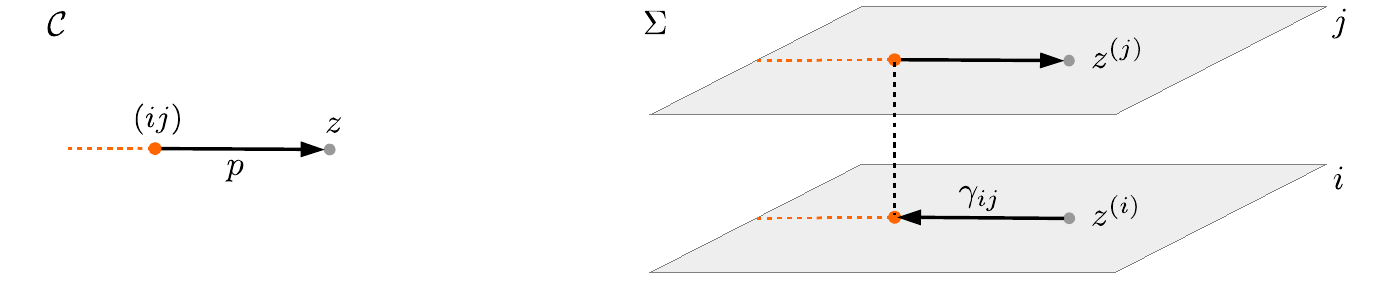}
\caption{\emph{Left:} Soliton path $p$ from a branch point of type $(ij)$ to a point $z$ on $\cC$, where a surface defect $\bS_z$ is located. \emph{Right:} Lift of $p$ to a 1-chain $\gamma_{ij}$ from $z^{(i)}$ to $z^{(j)}$ on $\Sigma$.}
\label{soliton}
\end{figure}

There is another kind of BPS states, called \emph{solitons} (or 2d-4d BPS states), that are bound to half-BPS surface defects $\bS_z$ in the 4d theory.
A surface defect $\bS_z$ is parameterized by a point $z$ on $\cC$, and generically has $K$ distinct massive vacua, which correspond to the $K$ lifts of $z^{(i)}$ to the sheets of $\Sigma$.
Solitons interpolate between two distinct vacua $z^{(i)}$ and $z^{(j)}$.
They are realized geometrically as open webs of strings on $\cC$ in which one of the strings ends at $z$.
Their charges are given by relative homology classes of 1-chains $\gamma_{ij}$ on $\Sigma$ with endpoints $z^{(i)}$ and $z^{(j)}$.
The basic example is a string extending along an open path $p$ between a branch point and $z$ (figure~\ref{soliton}).
The central charge and the mass are again given by~\eqref{ZMlambda}, from which it follows that BPS solitons correspond to $ij$-trajectories satisfying~\eqref{ijTraj}, for some fixed phase $\vartheta$.

A spectral network $\cW_\vartheta$ with phase $\vartheta$ is a representation on $\cC$ of all BPS solitons of phase $\vartheta$, that is a collection of $ij$-trajectories~\eqref{ijTraj}, for all ordered pairs $ij$:
\bea
\cW_\vartheta = \bigcup_{ij} \ \{ \text{$ij$-trajectories supporting a soliton of phase $\vartheta$} \}~.
\eea
This gives a network on $\cC$, made out of walls that start at branch points and asymptote to punctures (walls can also appear or disappear at junctions).
Letting $z$ be a local coordinate with $z=0$ at a branch point, we have $Z \simeq z^{3/2}$, which implies that there are three walls emerging from the branch point. 
Note that for generic values of $\vartheta$, the network $\cW_\vartheta$ does not contain any closed web of strings that correspond to 4d BPS states. This happens however for critical values $\vartheta_c$ for which an $ij$-wall and a $ji$-wall collide, and the topology of $\cW_\vartheta$ changes.

There is one more type of BPS states that plays an important role, namely \emph{framed BPS states}.
An open path $\wp$ from $z_1$ to $z_2$ on $\cC$ determines a pair of surface defects $\bS_{z_1}$ and $\bS_{z_2}$ as well as a supersymmetric interface $L_\wp$ between them.
The line defect $L_\wp$ should only depend on the homotopy class of $\wp$. 
The framed 2d-4d BPS states are the supersymmetric states of this combined system.
Geometrically, a framed 2d-4d BPS state is represented by a path $\gamma$ on $\Sigma$ which is essentially a lift of $\wp$ from $z_1^{(i)}$ to $z_2^{(j)}$.
More precisely, at each point where $\wp$ intersects a wall of the spectral network, the path $\gamma$ can make a \emph{detour} along a soliton path $\gamma_{ij}$ from the intersection point to the branch point and back. 
In such a case, the lifts of the segments of $\wp$ before and after the wall are on different sheets.
The projection of $\gamma$ to $\cC$ was compared to a millipede in~\cite{Gaiotto:2010be}, with the body corresponding to $\wp$, and the legs to the detours.

The case where $\wp$ is a closed path and $\bS_{z_1}=\bS_{z_2}$ are the null surface defect at $z_1=z_2$ corresponds to the line defect $L_\wp$ discussed in the introduction.
The framed BPS states supported on $L_\wp$ are associated with paths $\gamma$ on $\Sigma$ that are lifts of $\wp$ with possible detours along the walls of $\cW_\vartheta$ intersected by $\wp$.

%%%%%%%%%%%%%%
\subsection{Fock-Goncharov spectral networks} \label{secFGnetworks}

\begin{figure}[tb]
\centering
\includegraphics[width=\textwidth]{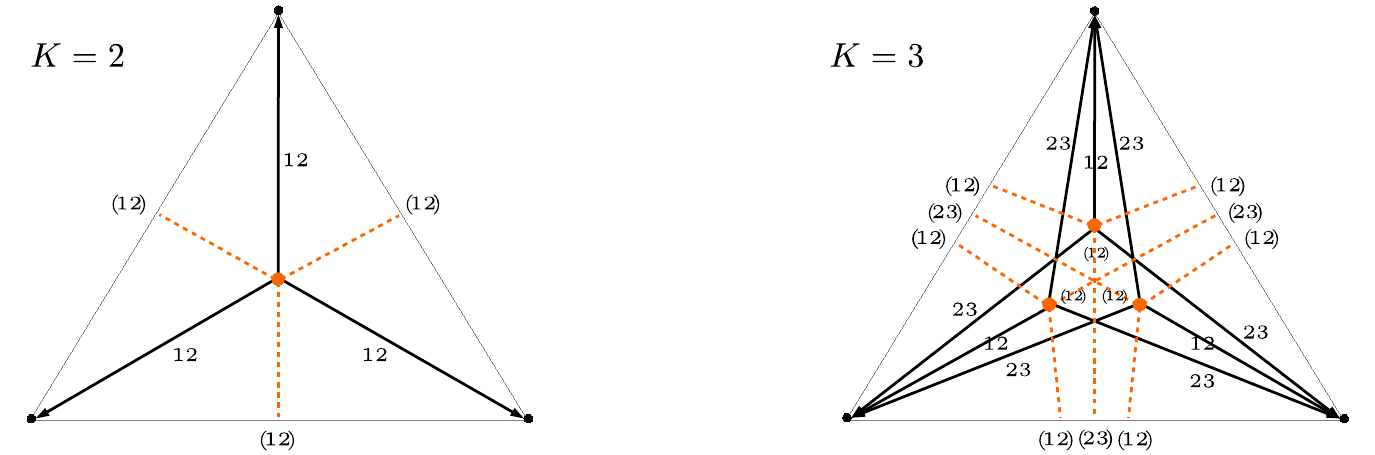}
\caption{Examples of Fock-Goncharov spectral networks $\cW$ on a triangle $\cC$ for $K=2$ and $K=3$. The walls (black arrows) of $\cW$ begin at branch points (orange dots) of the $K$-fold covering $\Sigma\to \cC$, and end at the marked points (black dots) on the boundary. They are labeled by ordered pairs $ij$ of sheets of $\Sigma$. Branch cuts (dashed orange lines) are labeled by pairs $(ij)$.}
\label{FGSpecNet}
\end{figure}

While general spectral networks for $K>2$ can be very complicated, an interesting family of tractable spectral networks was defined in~\cite{Gaiotto:2012db}.
Their advantage is that they are closely related to spectral networks for $K=2$, which are simply dual to ideal triangulations of~$\cC$ (with vertices at marked points). This provides a direct connection with the work of Fock and Goncharov on higher Teichm\"uller theory~\cite{FG}.
For simplicity, we focus on this family of \emph{Fock-Goncharov spectral networks} in this paper.
This will provide us with a nice way to organize changes of relative elevation along $[0,1]$ in section~\ref{secQHol}.
We leave the extension of our approach to more general spectral networks for future work.

The idea used in~\cite{Gaiotto:2012db} is to start with a Seiberg-Witten curve~\eqref{SigmaCurve} for $K=2$ and apply to it a homomorphism $\rho: SU(2) \to SU(K)$ given by the $K$-dimensional irreducible representation of $SU(2)$. The resulting curve is reducible, but can be made irreducible by a small perturbation. As an example, for $K=3$, starting with $\lambda^2 + \phi_2 = 0$ we obtain a curve of the form
\bea
\lambda^3 + (2\phi_2 + \delta \phi_2) \lambda + \delta \phi_3 = 0~.
\eea
Each of the branch points of the original curve splits into three slightly separated branch points for the new curve. Far away from these branch points, the walls of the associated spectral network align closely with the walls of the $K=2$ spectral network.
We show this procedure in figure~\ref{FGSpecNet} for the case where $\cC$ is simply a triangle (a disc with three marked points on the boundary).

Although no branch cut is necessary in the case of a triangle (since it is contractible), we choose a symmetric arrangement of three branch cuts at each branch point, which will be convenient when gluing triangles to obtain more elaborate Riemann surfaces $\cC$.
With this choice of branch cuts, all the branch points are of type $(12)$.
Some triplets of branch cuts of type $(i,i+1)$ meet at points in the middle of the triangle and continue as branch cuts of type $(i+1,i+2)$. This pattern is motivated by the requirement that the total transformation when crossing all the branch cuts along an edge of the triangle be the permutation $\{ 1,2, \ldots, K-1,K\} \to \{K, K-1, \ldots, 2,1\}$.

For general $K$, each branch point of the 2-fold covering splits into $\frac12 K(K-1)$ nearby branch points of the $K$-fold covering. The corresponding spectral network has three walls emerging from each branch point and ending at punctures. A collection of $\frac12 K(K-1)$ walls that parallel a wall of the $K=2$ spectral network is referred to as a \emph{cable}. The labels $ij$ of the walls in each cable can be chosen such that there is one wall with label $12$, two with $23$, three with $34$, and so on (this was called a minimal spectral network of \emph{Yang} type in~\cite{Gaiotto:2012db}).
This choice ensures that no new wall is created at any of the intersections.

%%%%%%%%%%%%%%
%%%%%%%%%%%%%%
\section{Nonabelianization} \label{secNonab}

Spectral networks lead to a process called \emph{nonabelianization}~\cite{Gaiotto:2012rg}, by which flat $GL(K,\bC)$-connections on a Riemann surface $\cC$ are described in terms of flat abelian connections on a $K$-fold branched cover $\Sigma$ of $\cC$.
Nonabelianization provides coordinates on the moduli space of flat connections, which are closely related to the Fock-Goncharov coordinates and admit a natural quantization.

%%%%%%%%%%%%%%
\subsection{From abelian to nonabelian flat connections} \label{ab2nonab}

Given a spectral network $\cW$ subordinate to a $K$-fold covering $\Sigma\to \cC$, the nonabelianization map $\Psi_\cW$ is defined as follows.
Consider a flat abelian connection $\nab$ in a line bundle $\cL$ over $\Sigma'$, where the prime indicates that the branch points of the covering $\pi : \Sigma \to \cC$ are removed. The push-forward $\pi_*(\nab)$ gives a flat connection in the rank-$K$ vector bundle 
\bea
E' = \pi_*(\cL)
\eea
over $\cC'$.
However, $\pi_*(\nab)$ cannot be extended to a flat connection on all of $\cC$ because it has non-trivial monodromies around the branch points.
These monodromies can be eliminated by cutting $\cC'$ along the walls of the spectral network $\cW$ and  regluing the connection with certain (non-diagonal) transition functions.
This produces the desired flat $GL(K,\bC)$-connection 
\bea
\nabla = \Psi_\cW(\nab)
\eea 
in a rank-$K$ vector bundle $E$ over $\cC$.

\begin{figure}[tb]
\centering
\includegraphics[width=\textwidth]{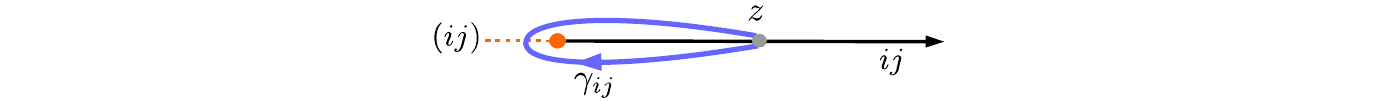}
\caption{Detour $\gamma_{ij}$ along an $ij$-wall, starting at $z^{(i)}$ on sheet $i$ and ending at $z^{(j)}$ on sheet $j$.}
\label{detourz}
\end{figure}

A little more precisely (see section 10 of~\cite{Gaiotto:2012rg} for full detail), in each component of $\cC'\backslash \cW$ the rank-$K$ vector bundle decomposes as 
\bea
E' = \bigoplus_{i=1}^K \cL_i~,
\eea
with $\cL_i$ denoting the restriction of the line bundle $\cL$ to sheet $i$, and $\pi_*(\nab)$ is diagonal.
In each component we have simply $\nabla \simeq \pi_*(\nab)$, and we can construct a basis of flat sections $s=(s_1, s_2, \ldots, s_K)$ solving $\nabla s = 0$.
The transition function between two components separated by an $ij$-wall $w\subset \cW$ is given by
\bea \label{transitionFct}
\cT_{ij} = 1 + \cX_{\gamma_{ij}} ~,
\eea
where $ \cX_\gamma $ denotes the parallel transport of $\nab$ along the path $\gamma$.
The path $\gamma_{ij}$ appearing in~\eqref{transitionFct} is a \emph{detour} along the $ij$-wall $w$, that is a path that starts at the lift $z^{(i)}$ to sheet $i$ of a point $z$ on $w$ (any point), circles the branch point at the origin of $w$, and comes back to the lift $z^{(j)}$ on sheet $j$ (figure~\ref{detourz}).
As described in section~\ref{secSpecNetN2Th}, detours correspond to BPS solitons (figure~\ref{soliton}).
The bundle $E$ is obtained by gluing together the restrictions of $E'$ to the components of $\cC'\backslash \cW$ with the transformations $\cT_{ij}$ along the walls of $\cW$.

We can be more explicit by choosing arbitrary flat sections $s_k^A \in \cL^A_k$ in a component $A$ and $s_k^B \in \cL^B_k$ in a component $B$, with $k=1, \ldots ,K$. The transition function~\eqref{transitionFct} relates these sections at any point $z$ on the $ij$-wall $w$. We note first that when $k\neq i$ we simply have $s_k^B =s_k^A$. In contrast, when $k=i$ we have
\bea
s_i^B = s_i^A+ s_i^A\cX_{\gamma_{ij}} ~.
\eea
Given that $s_i^A\cX_{\gamma_{ij}}  \in \cL^A_j$, there must be a constant $\kappa$ such that 
\bea \label{kappasjB}
s_i^A\cX_{\gamma_{ij}}  = \kappa s_j^A~.
\eea
The transformation $\cT_{ij}$ thus corresponds to an upper-triangular matrix:
\bea \label{transfoUpperTriang}
\begin{pmatrix} s_1^A \\ \vdots \\ s_i^A \\ s_j^A \\ \vdots \\ s_K^A \end{pmatrix} 
\quad \overset{\cT_{ij}}\mapsto \quad   \begin{pmatrix} s_1^B \\ \vdots \\ s_i^B \\ s_j^B \\ \vdots \\ s_K^B \end{pmatrix} =  
\begin{pmatrix} 1   &    &  &  & &    \\  & \; \ddots \; &   &  & &    \\    &   & 1 & \; \kappa   &    &   \\   &   & 0 &  \;1  & &  \\   &   &   &  & \; \ddots \; &  \\   &    &   &  & &   1 \end{pmatrix} 
\begin{pmatrix} s_1^A \\ \vdots \\ s_i^A \\ s_j^A \\ \vdots \\ s_K^A \end{pmatrix} ~.
\eea
This matrix can be expressed compactly as 
\bea \label{TijSL2embed}
\cT_{ij} = \varphi_i \cdot \begin{pmatrix}1& \kappa \\ 0&1
\end{pmatrix}  ~,
\eea 
where $\varphi_i : SL(2,\bC) \to GL(K,\bC)$ is the canonical embedding corresponding to the $i$\th root.

Nonabelianization actually produces a bit of extra structure around the punctures. 
In the simplest case of a puncture with walls of all possible types, the total gluing transformation is upper triangular, and thus preserves the subspaces
\bea
F^i = \bigoplus_{j> i} \cL_j~,
\eea
for $i=0,1,\ldots,K$.
This structure can be expressed as a $\nabla$-invariant \emph{flag}:
\bea
0 = F^K \subset F^{K-1} \subset F^{K-2} \subset \cdots \subset F^0 =   E~,
\eea
where $F^i$ has dimension $K-i$. 
This implies that $\nabla$ is a \emph{framed} flat $GL(K,\bC)$-connection on $\cC$.

We remark that in physical applications of spectral networks the relevant flat connections are in $SL(K,\bC)$ rather than $GL(K,\bC)$ (slightly oversimplifying, see~\cite{Gaiotto:2010be}).
Moreover, in~\cite{Gaiotto:2010be, Gaiotto:2012rg} the flat connections are actually \emph{twisted}, in the sense that they live in the unit tangent bundle over $\cC$, such that the holonomy around each $S^1$ fiber is $-1$. We will not address these subtleties in this paper, and content ourselves with flat $PGL(K,\bC)$-connections.

%%%%%%%%%%%%%%
\subsection{Coordinates on moduli spaces of flat connections} \label{CoordMflat}

Nonabelianization provides a map between the moduli space of flat abelian connections $\nab$ on $\Sigma$ and the moduli space of framed flat $GL(K,\bC)$-connections $\nabla$ on $\cC$:
\bea
\Psi_\cW : \quad \cM(\Sigma, GL(1,\bC)) \to \cM_F(\cC, GL(K,\bC))~.
\eea
In the case of a Riemann surface $\cC$ of genus $g$ with $n$ punctures and no boundary, the moduli space $\cM_F(\cC, GL(K,\bC))$ can be represented as the space of $GL(K,\bC)$-matrices $A_1, \ldots, A_g$, $B_1, \ldots, B_g$, and $M_1, \ldots, M_n$, subject to the relation
\bea
\prod_{a=1}^g A_a B_a A_a^{-1}B_a^{-1}  = \prod_{b=1}^n M_b~,
\eea
and considered up to overall conjugation. The holonomies $M_b$ around the punctures of $\cC$ have fixed eigenvalues. We then get 
\bea
\dim \cM(\cC, GL(K,\bC)) = (2 g +n)K^2 -2(K^2-1) - nK~.
\eea
On the other hand, the Riemann-Hurwitz formula gives
\bea
\dim \cM(\Sigma, GL(1,\bC)) = 2 g_\Sigma =  (2g-2)K+B+2~,
\eea
where $B$ is the number of branch points (assumed to be simple).
The dimensions of these two moduli spaces match when
\bea \label{numberBranchPts}
B = (2g + n -2)K(K-1)~,
\eea
in which case the nonabelianization map $\Psi_\cW$ is one-to-one and can be understood as providing coordinate systems on $\cM_F(\cC, GL(K,\bC))$.
Condition~\eqref{numberBranchPts} does hold for the spectral curves~\eqref{SigmaCurve} in theories of class S. In particular, the spectral curves associated with ideal triangulations in section~\ref{secFGnetworks} have $\frac12 K(K-1)$ branch points in each of their $2(2g + n -2)$ triangles.

Coordinates on $\cM_F(\cC, GL(K,\bC))$ are given by holonomies of the flat abelian connection $\nab = \Psi_\cW^{-1}(\nabla)$:
\bea \label{parallelTranspAB}
\cX_\gamma = \Hol_\gamma \nab ~,
\eea
where $\gamma$ runs over a basis of $H_1(\Sigma'; \bZ)$ (see~\cite{Hollands:2013qza} for a discussion of the \emph{abelianization map} $\Psi_\cW^{-1}$ for $SL(2,\bC)$). 
These coordinates only dependent on the homotopy class of $\gamma$.
However, as mentioned in section~\ref{ab2nonab}, the flat abelian connection $\nab$ cannot be extended smoothly across the branch points. 
Instead, $\nab$ picks up a minus sign as it moves through a branch point on $\Sigma$ (such a connection was called ``almost-flat'' in~\cite{Hollands:2013qza}):
\bea \label{XgAlmostFlat}
\cX_\gamma = - \cX_{\gamma'}
\eea
for paths $\gamma$ and $\gamma'$ on different sides of a branch point.
The Poisson bracket is expressed in terms of the intersection pairing $\langle \cdot, \cdot \rangle$ on $H_1(\Sigma'; \bZ)$:
\bea
\{ \cX_{\gamma_1}, \cX_{\gamma_2} \} = \langle \gamma_1, \gamma_2 \rangle \cX_{\gamma_1\gamma_2}~.
\eea
The $\cX_{\gamma}$ consist of Darboux coordinates on the symplectic leaves of the moduli space together with central elements of the Poisson algebra.

\begin{figure}[tbh]
\centering
\includegraphics[width=\textwidth]{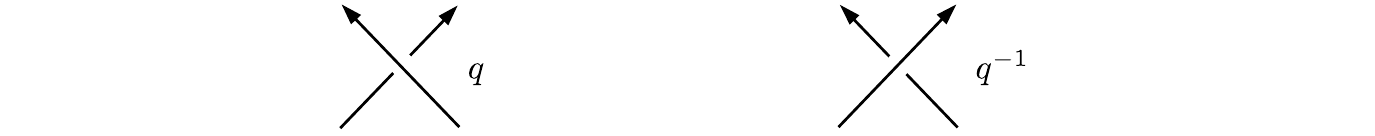}
\caption{A right- or left-handed intersection corresponds to $q$ or $q^{-1}$.}
\label{crossingsq}
\end{figure}

The coordinates $\cX_\gamma$ can be naturally quantized to noncommutative variables $\hat \cX_\gamma$ obeying the relation 
\bea \label{XgRelation}
\hat \cX_{\gamma_1} \hat \cX_{\gamma_2} = q^{\langle \gamma_1, \gamma_2 \rangle} \hat \cX_{\gamma_1\gamma_2} ~.
\eea
This means that a right-handed intersection corresponds to a factor of $q$, and a left-handed one to $q^{-1}$ (figure~\ref{crossingsq}).

\begin{figure}[tbh]
\centering
\includegraphics[width=\textwidth]{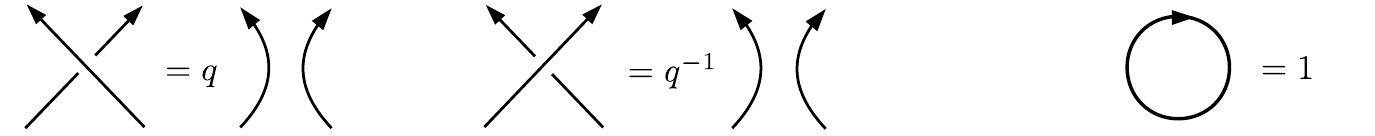}
\caption{Abelian skein relations.}
\label{abelianSkein}
\end{figure}

This can be thought of as a manifestation of the abelian skein relations.
A right- or left-intersections is simply resolved into non-intersecting paths with a coefficient of $q$ or $q^{-1}$, while a contractible loop can be deleted (figure~\ref{abelianSkein}).

\begin{figure}[tbh]
\centering
\includegraphics[width=\textwidth]{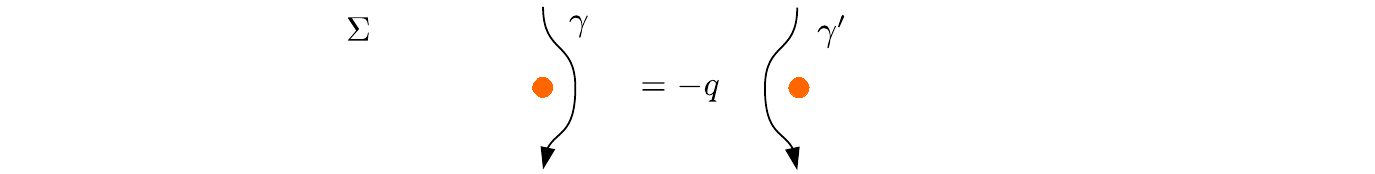}
\caption{The noncommutative variable $\hat \cX_\gamma$ picks up a factor of $-q$ when $\gamma$ is moved across a branch point on $\Sigma$.}
\label{almostFlatSigma}
\end{figure}

\begin{figure}[tbh]
\centering
\includegraphics[width=\textwidth]{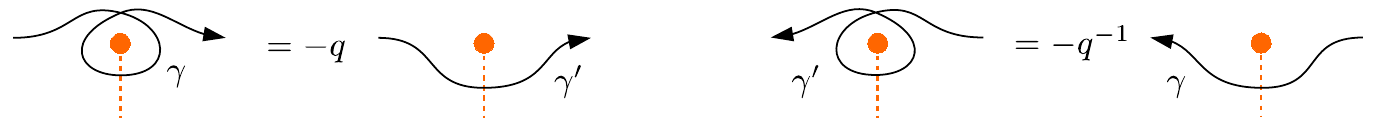}
\caption{When drawn on $\cC$, the move of figure~\ref{almostFlatSigma} looks like looping or unlooping a path around a branch point.}
\label{almostFlat}
\end{figure}

In analogy with~\eqref{XgAlmostFlat}, we impose that the noncommutative variables associated with paths $\gamma$ and $\gamma'$ on different sides of a branch point on $\Sigma$ (figure~\ref{almostFlatSigma}) are related by
\bea \label{XgAlmostFlatq}
\hat \cX_\gamma = -q \hat \cX_{\gamma'}~.
\eea
When drawn on $\cC$, the paths $\gamma$ and $\gamma'$ differ by a loop around the branch point (figure~\ref{almostFlat}).
This reflects the fact that $\nab$ has non-trivial monodromies around the branch points. 
We will see in section~\ref{secHomotopy} that the condition~\eqref{XgAlmostFlatq} leads to homotopy invariance of the quantum parallel transport.

%%%%%%%%%%%%%%
\subsection{Relation to Fock-Goncharov coordinates} \label{secFGcoords}

\begin{figure}[tbh]
\centering
\includegraphics[width=\textwidth]{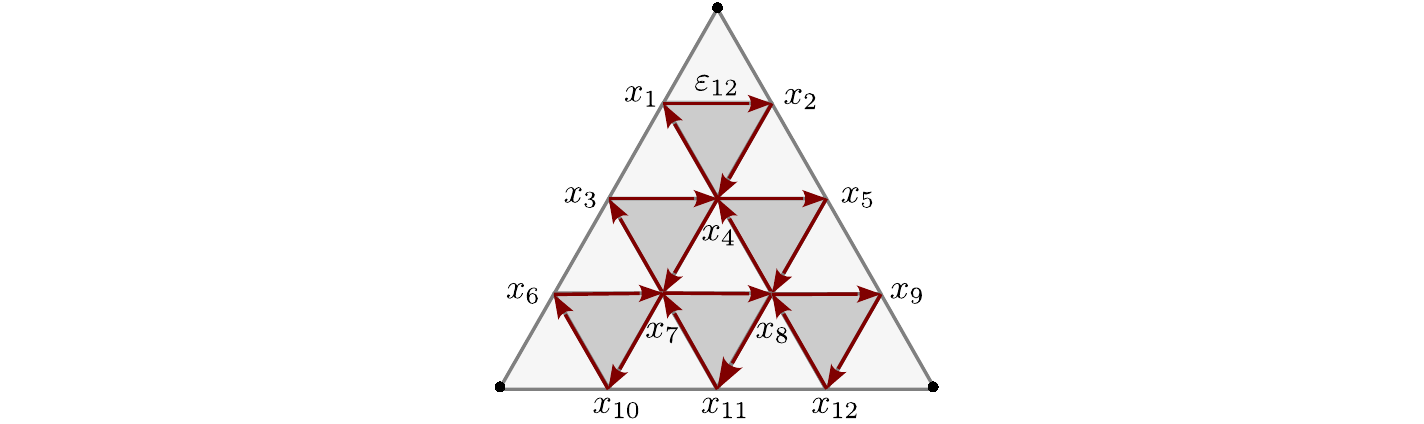}
\caption{$K$-triangulation of an ideal triangle into $K^2$ small black and white triangles (here for $K=4$). The Poisson structure $\varepsilon$ is encoded in the arrows circulating clockwise around the small black triangles.}
\label{Poisson}
\end{figure}

Fock and Goncharov defined useful systems of coordinates for $\cM_F(\cC, GL(K,\bC))$ associated with ideal triangulations of $\cC$~\cite{FG}. 
Each ideal triangle is itself decomposed into $K^2$ small triangles, which produces a so-called $K$-triangulation (figure~\ref{Poisson}).
The Fock-Goncharov coordinates $x_\alpha$ are associated with the vertices of these small triangles (excluding the punctures of $\cC$).
There are $K-1$ coordinates on each edge, and $\frac12 (K-1)(K-2)$ coordinates inside each face.

The Poisson structure is encoded in a system of oriented arrows on the edges of the small triangles of the $K$-triangulation:
\bea\label{PoissonFG}
\{ x_\alpha, x_\beta \} = \varepsilon_{\alpha\beta} x_\alpha x_\beta ~,
\eea
with
\bea\label{PoissonTensorFG}
\varepsilon_{\alpha\beta}  = \#(\text{arrows from $x_\alpha$ to $x_\beta$}) - \#(\text{arrows from $x_\beta$ to $x_\alpha$})  ~.
\eea
The $x_\alpha$ can be naturally quantized to noncommutative variables $\hat x_\alpha$ satisfying
\bea
\hat  x_\alpha \hat x_\beta = q^{2\varepsilon_{\alpha\beta}} \hat x_\beta\hat x_\alpha~.
\eea
The logarithmic coordinates $X_\alpha$ defined via $x_\alpha = \exp X_\alpha$ quantize to noncommutative variables $\hat X_\alpha$ satisfying
\bea \label{logCoordX}
[\hat X_\alpha, \hat X_\beta] = 2 \hbar \{X_\alpha, X_\beta\} = 2 \hbar \varepsilon_{\alpha\beta}~,
\eea
where $q = \exp \hbar$. 

\begin{figure}[tbh]
\centering
\includegraphics[width=\textwidth]{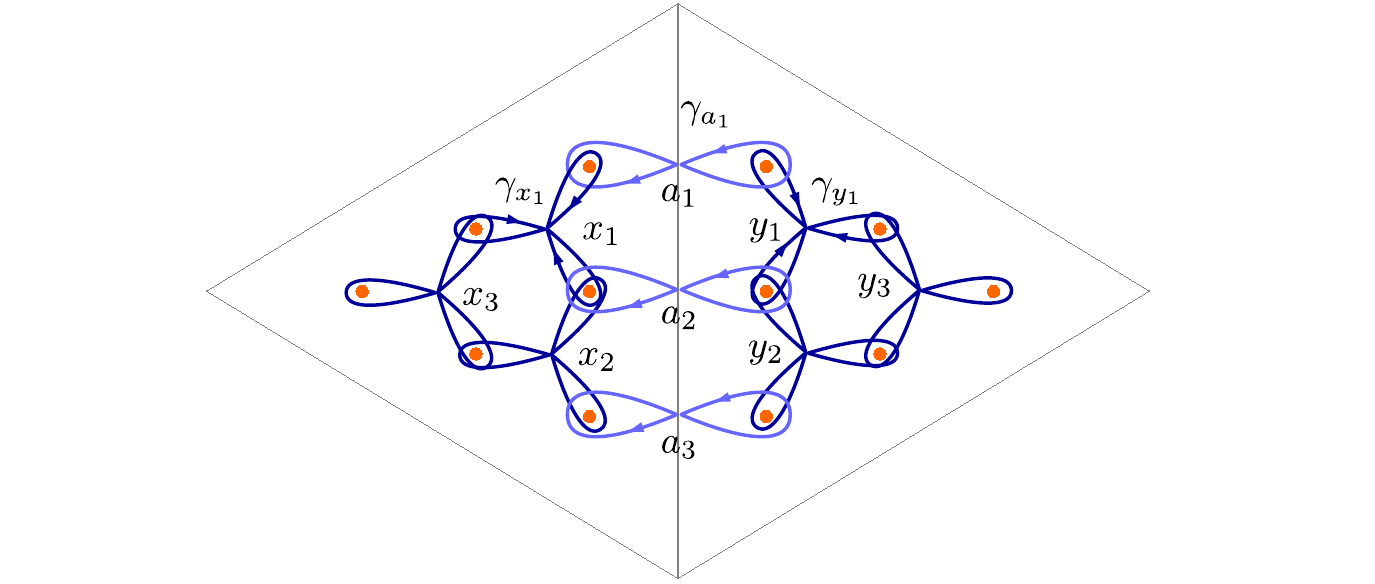}
\caption{Edge coordinates correspond to loops around pairs of branch points, face coordinates to loops around triplets of branch points (here for $K=4$).}
\label{FGloops}
\end{figure}

The Fock-Goncharov coordinates $x_\alpha$ can be identified with the coordinates $\cX_\gamma$ provided by nonabelianization, for certain choices of paths $\gamma$~\cite{Gaiotto:2010be, Hollands:2013qza}. 
A Fock-Goncharov coordinate~$x$ on an edge of an ideal triangulation of $\cC$ is a cross-ratio constructed from a quadruplet of vectors coming from the four flags at the vertices of the quadrilateral containing the edge.
It coincides with the abelian parallel transport along a loop $\gamma_x$ surrounding a pair of branch points, one on each side of the edge (figure~\ref{FGloops}):
\bea
\cX_{\gamma_x} = x~.
\eea
Similarly, a Fock-Goncharov coordinate inside the face of a triangle is a triple ratio and corresponds to a loop surrounding three branch points.
The intersection matrix of all these loops reproduces the Poisson tensor $\varepsilon$ in~\eqref{PoissonTensorFG}.

%%%%%%%%%%%%%%%%%%%
%%%%%%%%%%%%%%%%%%%
\section{Quantum parallel transport} \label{secQPT}

Spectral networks provide a construction of the parallel transport $F(\wp)$ of the $GL(K,\bC)$-connection $\nabla$ along an open path $\wp$ on $\cC$ in terms of abelian holonomies on the cover $\Sigma$~\cite{Gaiotto:2012rg}.
It is invariant under homotopy of $\wp$ and relates naturally to Fock and Goncharov's snake matrices.
The corresponding quantum parallel transport $F(\wp;q)$ was studied in~\cite{Galakhov:2014xba}.

%%%%%%%%%%%%%%
\subsection{Detours}

The parallel transport of $\nabla$ along an \emph{open} path $\wp$ from $z_1$ to $z_2$ on $\cC$ is defined as a sum over certain paths $\gamma$ from $z_1^{(i)}$ to $z_2^{(j)}$ on $\Sigma$, for all sheets $i$ and $j$:
\bea \label{genFunctionF}
F(\wp) = \sum_\gamma \underline{\overline{\Omega}}(\wp,\gamma) \cX_\gamma~.
\eea
Here $\cX_\gamma$ denotes the parallel transport of $\nab=\Psi_\cW^{-1}(\nabla)$ along $\gamma$ as in~\eqref{parallelTranspAB}, and the coefficients $\underline{\overline{\Omega}}(\wp,\gamma)$ are the framed BPS indices.
The paths $\gamma$ are lifts of $\wp$ to $\Sigma$ that can take all possible detours along walls of $\cW$ that $\wp$ intersects.
We have indeed seen that $\nabla$ is obtained by gluing along the walls of $\cW$ with the transition functions~\eqref{transitionFct}. For a path $\wp$ that intersects a single $ij$-wall, at which point it splits as $\wp = \wp_1\wp_2$, we can thus write
\bea \label{FpDpDetours}
F(\wp) =  \sum_{k=1}^K \cX_{\wp^{(k)}} +  \cX_{\wp_1^{(i)}}   \cX_{\gamma_{ij}} \cX_{\wp_2^{(j)}} ~,
\eea
where $\wp^{(k)}$ is the lift of $\wp$ to sheet $k$ of $\Sigma$.
The second term in~\eqref{FpDpDetours} involves the concatenated path $\wp_1^{(i)} \gamma_{ij} \wp_2^{(j)}$, that is the lift of $\wp_1$ to sheet $i$, followed by a detour $\gamma_{ij}$ around the branch point at the origin of the wall, and completed by the lift of $\wp_2$ to sheet $j$.
As mentioned in section~\ref{secSpecNetN2Th}, the detour $\gamma_{ij}$ corresponds to a BPS soliton interpolating between two vacua of a surface defect $\bS_z$.
For the case of a path $\wp$ on $\cC$ that intersects $\cW$ multiple times, $F(\wp)$ is defined by breaking it into subpaths that intersect $\cW$ once, and repeatedly using the composition property 
\bea
F(\wp_1\wp_2) = F(\wp_1) F(\wp_2) ~.
\eea
This gives a concrete recipe for computing $F(\wp)$.

The quantum parallel transport is defined by promoting the $\cX_\gamma$ to the noncommutative variables $\hat \cX_\gamma$:
\bea \label{NCparallelTransp}
F(\wp;q) = \sum_\gamma \underline{\overline{\Omega}}(\wp,\gamma;q) \hat \cX_\gamma~,
\eea
where the coefficients $\underline{\overline{\Omega}}(\wp,\gamma;q)$ are the framed protected spin characters~\eqref{framedPSCdef} (this agrees with the generating function~\eqref{GenFctPSC}).
It can again be computed by applying the detour rule at each intersection with a wall of $\cW$:
\bea \label{FpDpDetoursQ}
F(\wp;q) =  \sum_{k=1}^K \hat \cX_{\wp^{(k)}} +  \hat \cX_{\wp_1^{(i)}}   \hat \cX_{\gamma_{ij}} \hat \cX_{\wp_2^{(j)}}  ~.
\eea
Recall that the $\hat \cX_\gamma$ satisfy~\eqref{XgRelation}, which corresponds to the abelian skein relations of figure~\ref{abelianSkein}, and that moving $\gamma$ across a branch point produces a factor of $-q$, as in~\eqref{XgAlmostFlatq} (figures~\ref{almostFlat}).

%%%%%%%%%%%%%%%%%%%
\subsection{Homotopy invariance}\label{secHomotopy}

An important property of the quantum parallel transport $F(\wp;q)$ is that it is invariant under homotopy of $\wp$ on $\cC$, as we will now show
(following~\cite{Gaiotto:2012rg, Galakhov:2014xba}).
For any two paths $\wp$ and $\wp'$ related by homotopy we should have
\bea\label{homotopyF}
F(\wp;q) =F(\wp';q) ~.
\eea

\begin{figure}[tb]
\centering
\includegraphics[width=\textwidth]{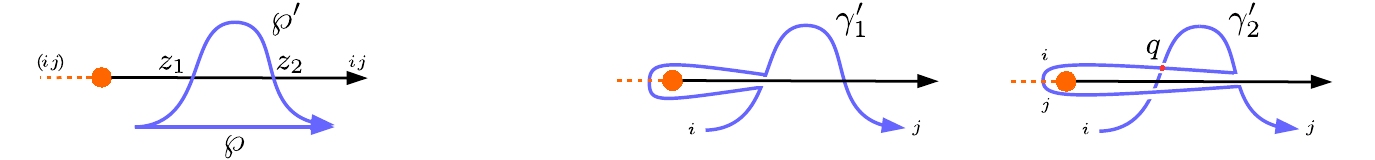}
\caption{\emph{Left}: Paths $\wp$ and $\wp'$ on $\cC$ related by a homotopy through an $ij$-wall.
\emph{Right}: Lifted paths $\gamma'_1$ and $\gamma'_2$ on $\Sigma$, with detours around the branch point of type $(ij)$. Note that $\gamma_2'$ has a right-handed self-intersection.}
\label{homotopyWall}
\end{figure}

\begin{figure}[htb]
\centering
\includegraphics[width=\textwidth]{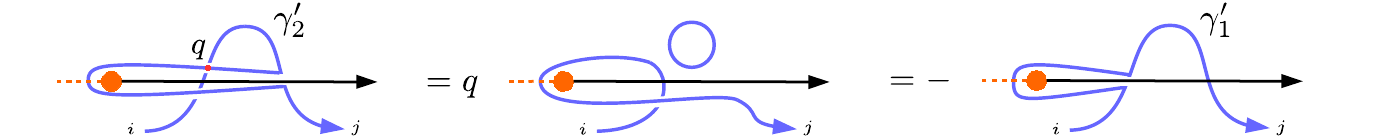}
\caption{The contributions from $\gamma_1'$ and $\gamma_2'$ on the right of figure~\ref{homotopyWall} cancel each other upon applying the abelian skein relations and moving across the branch point.}
\label{homotopyWall2}
\end{figure}

Let us consider the case where $\wp$ and $\wp'$ are related by a homotopy through an $ij$-wall (figure~\ref{homotopyWall}).
The path $\wp$ does not cross any wall and we just have
\bea
F(\wp;q) =   \sum_{k=1}^K \hat \cX_{\wp^{(k)}}  ~.
\eea
On the other hand, the path $\wp'$ crosses the wall twice, say at $z_1$ and $z_2$, so the parallel transport involves paths with a detour starting on sheet $i$ and ending on sheet $j$:
\bea
F(\wp';q) = \sum_{k=1}^K \hat \cX_{\wp'^{(k)}}   + \hat \cX_{\gamma_1'}   +  \hat \cX_{\gamma_2'}~,
\eea
where the detoured paths are $\gamma_1' = \wp_1'^{(i)} \gamma_{ij}(z_1) \wp_2'^{(j)} $ and $\gamma_2' = \tilde\wp_1'^{(i)}\tilde \gamma_{ij}(z_2) \tilde\wp_2'^{(j)} $.
The first term matches $F(\wp;q)$ since the abelian parallel transports $\cX_\gamma$ and their quantizations $\hat \cX_\gamma$ only depend on the homotopy class of $\gamma$. 
Homotopy invariance of $F(\wp;q)$ therefore requires that 
the contributions from the two paths $\gamma'_1  $ and $\gamma'_2 $ cancel each other.
To see this, we first apply the abelian skein relation to resolve the intersection of $\gamma_2'$, which produces a factor of $q$ (figure~\ref{homotopyWall2}).
We obtain a contractible loop and a path that circles the branch point counter-clockwise, which by the rule of figure~\ref{almostFlat} can be replaced by a straight path homotopic to $\gamma_1'$ with a factor of $-q^{-1}$.
The factors combine to give
\bea
\hat \cX_{\gamma_2'} = - \hat \cX_{\gamma_1'}~,
\eea
so that the homotopy invariance~\eqref{homotopyF} indeed holds.

\begin{figure}[htb]
\centering
\includegraphics[width=\textwidth]{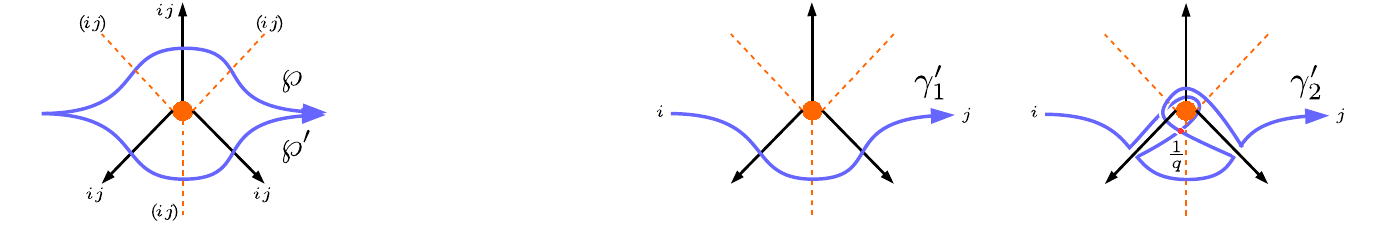}
\caption{\emph{Left}: Paths $\wp$ and $\wp'$ related by a homotopy across a branch point.
\emph{Right}: Lifted paths $\gamma'_1$ and $\gamma'_2$ that cancel each other.}
\label{homotopyPoint}
\end{figure}

Another interesting case is when $\wp$ and $\wp'$ differ by a homotopy across a branch point (figure~\ref{homotopyPoint}).
In fact, it was the \textit{raison d'\^etre} of the gluing with detour~\eqref{transitionFct} to ensure that the connection $\nabla$ extends smoothly across branch points.
We again find that a pair of undesired paths $\gamma'_1$ and $\gamma'_2$ lifted from $\wp'$ cancel out via a combination of factors $q^{-1}$ and $-q$ coming from a self-intersection and a loop around the branch point.

It is straightforward to see that a homotopy across a joint where two walls intersect also leave the quantum parallel transport $F(\wp;q)$ invariant.

%%%%%%%%%%%%%%
%%%%%%%%%%%%%%
\subsection{Relation to snake matrices}

We now relate the parallel transport across walls, edges, cables, or branch cuts to the \emph{snake matrices} used by Fock and Goncharov to construct $PGL(K,\bC)$ holonomies on a triangulated surface $\cC$~\cite{FG}.
The general procedure to obtain the holonomy for a curve $\wp$ is to decompose it into elementary matrices of two types corresponding to step-wise moves of a \emph{snake}, that is an oriented path in a triangle from a vertex to the opposite edge, through small triangles of the $(K-1)$-triangulation.

\begin{figure}[tbh]
\centering
\includegraphics[width=\textwidth]{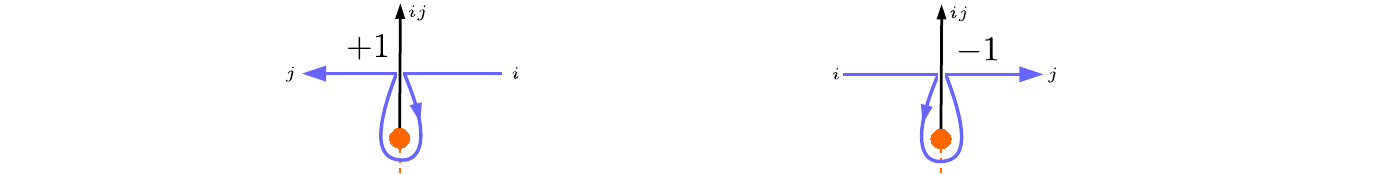}
\caption{Paths with detour along an $ij$-wall, corresponding to the off-diagonal components $\pm1$ in the elementary snake matrices $F_i$ and $F_i^{-1}$.}
\label{crossingInv}
\end{figure}

\paragraph{Across walls:}
The first elementary snake matrix corresponds to moving through a small black triangle clockwise and takes the form
\bea \label{snakeFi}
F_i = \varphi_i \begin{pmatrix} 1&1\\ 0&1\end{pmatrix} ~.
\eea
This agrees with the transition function~\eqref{TijSL2embed} across an $ij$-wall, provided that we fix the relative normalizations of 
$s_i^A$ and $s_j^A$ in~\eqref{kappasjB} such that $\kappa = 1$.
The inverse of $F_i$, corresponding to moving counter-clockwise, is 
\bea \label{snakeFiInv}
F_i^{-1} = \varphi_i \begin{pmatrix} 1&-1\\ 0&1\end{pmatrix} ~,
\eea
and requires setting
$\kappa = -1$.
This is summarized graphically in figure~\ref{crossingInv}.

%%%%%%%%%%%%%%
\paragraph{Across edges:}

The second elementary snake matrix is a diagonal matrix depending on a coordinate $x$:
\bea 
H_i (x) = \text{diag} (\underbrace{1, \cdots,1}_{i \text{ times}}, x, \cdots, x)~.
\eea
This matrix appears in particular when a snake moves across an edge of the triangulation. 
For an edge with coordinates $\{ x_1, x_2, \ldots, x_{K-1} \}$, in this order along the snake, the transformation is given by 
\bea \label{edgeSnakeMatrix}
M_E(x_1,x_2, \ldots, x_{K-1}) &=& H_1(x_1) H_2(x_2) \cdots H_{K-1}(x_{K-1}) \nn
&=&   \begin{pmatrix} 1 &&&&  \\ &  x_1 &&& \\ &&  x_1 x_2 &&\\  &&& \ddots& \\  &&&& x_1x_2 \cdots x_{K-1}  \end{pmatrix}~.
\eea

\begin{figure}[tbh]
\centering
\includegraphics[width=\textwidth]{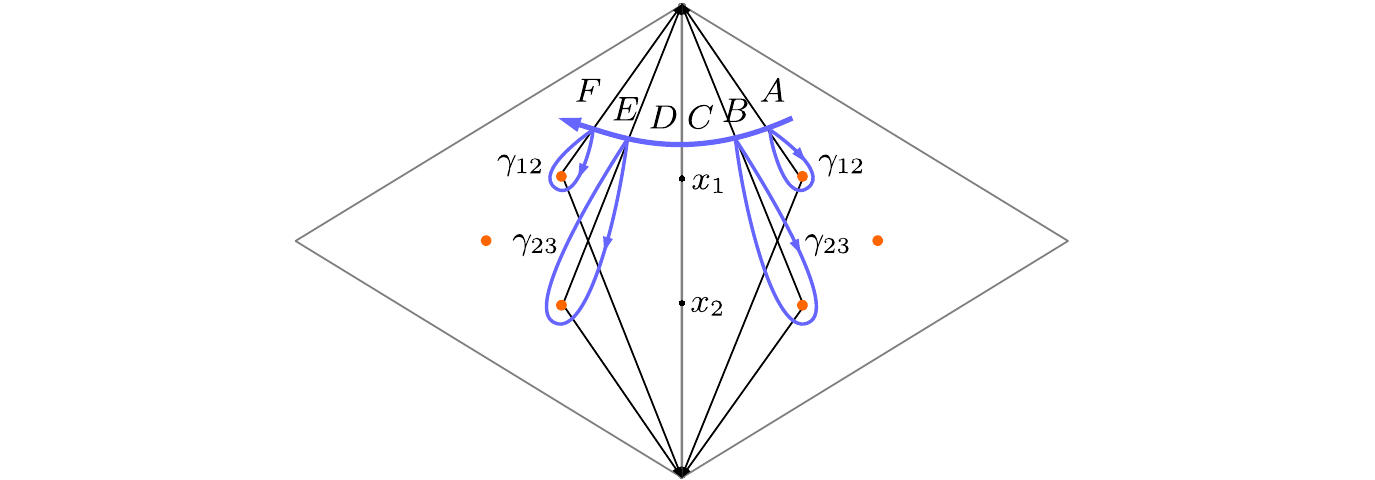}
\caption{The diagonal transition matrix across an edge of the triangulation can be determined by comparing various parallel transports with detours (branch cuts and irrelevant parts of the spectral network are omitted).}
\label{edgeTransition}
\end{figure}

To understand how this relates to the parallel transport $F(\wp)$, recall that in each component of $\cC\backslash \cW$ we are free to make a diagonal gauge transformation on the flat sections.
Let us consider a component that contains an edge of the triangulation, and divide it along the edge into two regions $C$ and $D$. We apply a diagonal gauge transformation such that the flat sections in region $D$ are proportional to the ones in region $C$:
\bea
s_i^D = \alpha_i s_i^C~.
\eea
We are going to show that the $\alpha_i$ are indeed products of Fock-Goncharov coordinates along the edge, as in~\eqref{edgeSnakeMatrix}.
We focus on the case $K=3$ and compare parallel transports along paths from sheet~1 in region $A$ to sheet~2 in region $F$ (figure~\ref{edgeTransition}). 
There are two such paths with a detour along a 12-wall, one between regions $A$ and $B$, and the other between regions $E$ and $F$. 
We know from~\eqref{kappasjB} with $\kappa=1$ that across these walls we have
\bea
s_1^A \cX_{\gamma_{12}^{AB}}  = s_2^B~, \qqq
s_1^E \cX_{\gamma_{12}^{EF}} =  s_2^F~.
\eea
Completing these relations to parallel transports from region $A$ to region $F$ we obtain
\bea
s_1^A \cX_{\gamma_{12}^{AB}}  =  \alpha_2^{-1} s_2^F~, \qqq
\alpha_1 s_1^A \cX_{\gamma_{12}^{EF}} =  s_2^F~,
\eea
where $\alpha_1$ and $\alpha_2$ appeared when crossing the edge on sheet 1 and 2.
We then find
\bea
\frac{\alpha_2 }{\alpha_1} = \frac{\cX_{\gamma_{12}^{EF}} }{ \cX_{\gamma_{12}^{AB}}} =  \cX_{\gamma_{12}^{EF}(\gamma_{12}^{AB})^{-1}} = x_1~.
\eea
In the last equality we used the fact that the path $\gamma_{12}^{EF}(\gamma_{12}^{AB})^{-1}$ is homotopic to the loop $\gamma_{x_1}$ that corresponds to the edge coordinate $x_1$ (section~\ref{secFGcoords}).
A similar calculation comparing paths from sheet 2 in $A$ to sheet 3 in $F$ gives
\bea
\frac{\alpha_3 }{\alpha_2} = \cX_{\gamma_{23}^{DE}(\gamma_{23}^{BC})^{-1}} = x_2~.
\eea
Setting $\alpha_1=1$, we conclude that the transformation across the edge takes the form
\bea
\cT (x_1,x_2) = \begin{pmatrix} 1 &&  \\ &  x_1 & \\ &&  x_1 x_2   \end{pmatrix}~,
\eea
in agreement with the snake matrix~\eqref{edgeSnakeMatrix} for $K=3$.

\begin{figure}[tbh]
\centering
\includegraphics[width=\textwidth]{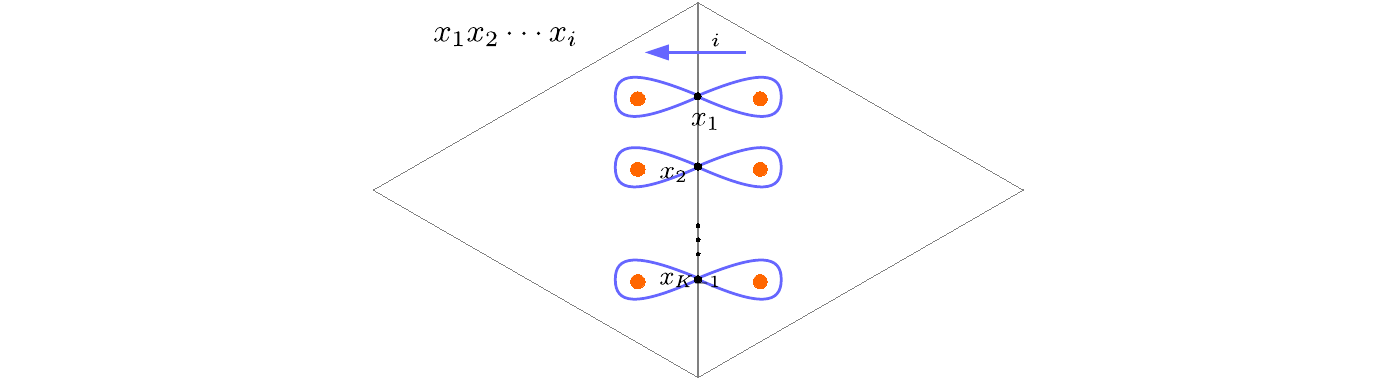}
\caption{A path on sheet~$i$ that crosses an edge with coordinates $\{x_1, \ldots, x_{K-1} \}$ picks up the product $x_1x_2\cdots x_i$.}
\label{edgeTransfo}
\end{figure}

In general, a path on sheet~$i$ that crosses an edge with coordinates $\{ x_1, x_2, \ldots, x_{K-1} \}$ (in this order) picks up the $i$\th eigenvalue of the matrix~\eqref{edgeSnakeMatrix} (figure~\ref{edgeTransfo}).
This matrix is defined close to the extremity of the edge near $x_1$, and crossing branch cuts will of course permute its eigenvalues.
Moving the path from one extremity of the edge to the other extremity exchanges all the eigenvalues, which in $PGL(K,\bC)$  is the same as taking the inverse:
\bea  \label{MEinv}
M_E^{-1}(x_1, x_2, \ldots, x_{K-1}) 
\simeq \begin{pmatrix} x_1x_2 \cdots x_{K-1} &&&&  \\ & \ddots&&& \\ &&  x_1 x_2 &&\\  &&&x_1& \\  &&&& 1  \end{pmatrix} ~.
\eea

%%%%%%%%%%%%%%
\paragraph{Across cables:}

An important transformation consists of moving a snake through a face of a triangle, from one edge to next. For a clockwise rotation, the corresponding snake matrix for $K=3$ is 
\bea \label{MFace}
M_F (x)= F_2F_1 H_2(x) F_2 = \begin{pmatrix}  1&1&1 \\ 0&1&1+x \\ 0&0&x \end{pmatrix}~,
\eea
with $x$ the face coordinate, while for a counter-clockwise rotation it is
\bea
M_F^{-1}(x) = \begin{pmatrix}  1&-1&1 \\ 0&1&-1-x^{-1} \\ 0&0&x^{-1} \end{pmatrix}    \simeq \begin{pmatrix}  x&-x&x \\ 0&x&-1-x \\ 0&0&1 \end{pmatrix}~.
\eea

\begin{figure}[tbh]
\centering
\includegraphics[width=\textwidth]{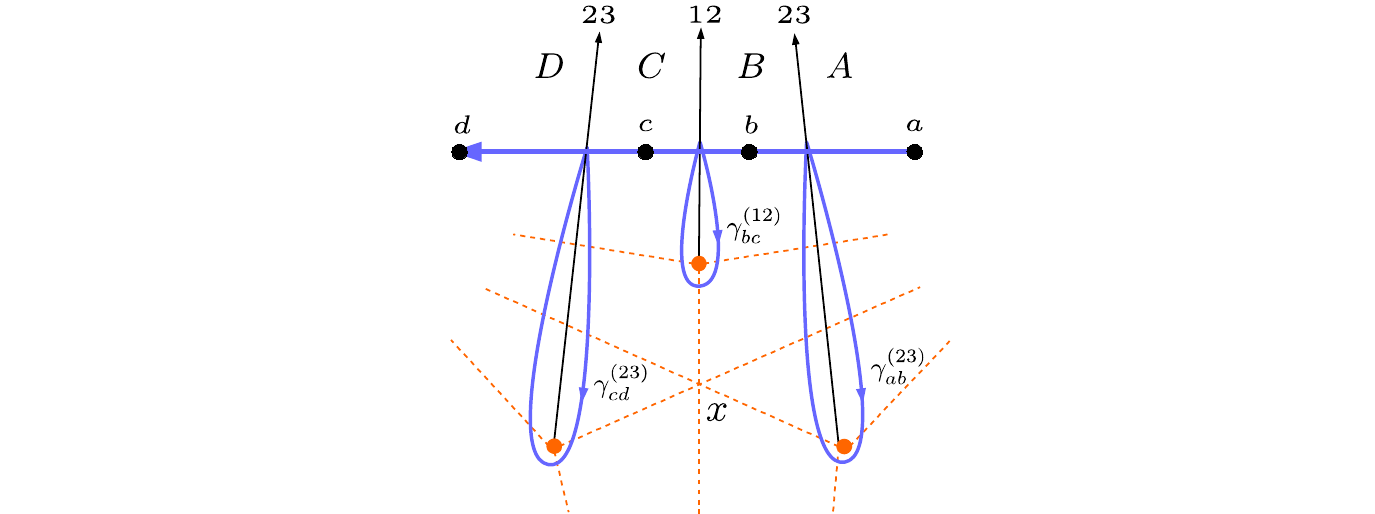}
\caption{Parallel transport across the three walls of a cable for $K=3$.}
\label{cableTransfo}
\end{figure}

\begin{figure}[tb]
\centering
\includegraphics[width=\textwidth]{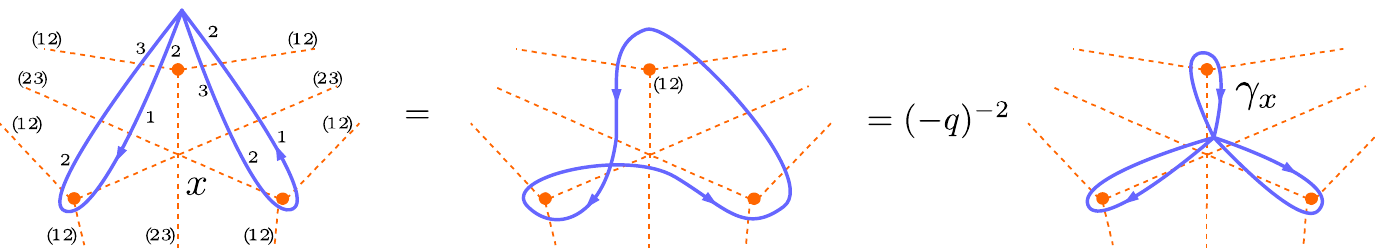}
\caption{The difference of the detours $\gamma^{(23)}_{cd}$ and $ \gamma^{(23)}_{ab}$ is equivalent, upon moving the segment on sheet~3 through the top branch point of type $(12)$ and using the relations in figure~\ref{almostFlat}, to the path $\gamma_x$ corresponding to the face coordinate $x$.}
\label{gammax}
\end{figure}

We will now show how to match $M_F (x)$ with the transformation for crossing the three walls of a cable (see section~6 in~\cite{Gaiotto:2012db}).
It is already clear from our previous discussion that the snake matrices $F_2$ in~\eqref{MFace} correspond to the transformations across the 23-walls between regions $A$ and $B$, and between regions $C$ and $D$ (figure~\ref{cableTransfo}).
To describe the parallel transport from region $B$ to region $C$ across the 12-wall, we define the following reference flat sections at the point $c$:
\bea  \label{refSectionsC}
s_i^C(c) &=& s_i^B(b) \cX_{\gamma^{(i)}_{bc}}~, \qqq i = 1,2~, \nn
s_3^C(c)&=& \alpha s_3^B(b) \cX_{\gamma^{(3)}_{bc}}~,
\eea
where $\cX_{\gamma^{(i)}_{bc}}$ is the abelian parallel transport along the path $\gamma^{(i)}_{bc}$ from $b$ to $c$ on sheet $i$.
The nonabelian parallel transport along a path $\wp_{bc}$ on $\cC$ from $b$ to $c$ is performed with~\eqref{genFunctionF}:
\bea
s_i^B(c) = s_i^B(b) F(\wp_{bc})~.
\eea
The detour rule~\eqref{FpDpDetours} gives
\bea
s_1^B(c) &=& s_1^B(b) \cX_{\gamma^{(1)}_{bc}}+s_1^B(b) \cX_{\gamma^{(12)}_{bc}}   ~, \nn
s_2^B(c) &=& s_2^B(b) \cX_{\gamma^{(2)}_{bc}}~, \nn
s_3^B(c) &=& s_3^B(b) \cX_{\gamma^{(3)}_{bc}}~.
\eea
Comparing with~\eqref{refSectionsC} and setting $s_1^B(b) \cX_{\gamma^{(12)}_{bc}}   = - s_2^B(c)$, we arrive at the transformation
\bea  
\begin{pmatrix}  s_1^C \\ s_2^C \\ s_3^C \end{pmatrix} =  \begin{pmatrix}  1& 1 & 0 \\ 0&1&0\\ 0&0& \alpha  \end{pmatrix} \begin{pmatrix}  s_1^B \\ s_2^B \\ s_3^B \end{pmatrix} = F_1 H_2(\alpha)  \begin{pmatrix}  s_1^B \\ s_2^B \\ s_3^B \end{pmatrix}   ~.
\eea
To determine $\alpha$ we compare the abelian parallel transports along two paths from $a$ to $d$ with a detour $\gamma^{(23)}_{ab}$ or $\gamma^{(23)}_{cd}$. We find
\bea
\alpha s_2^A(a) \cX_{\gamma^{(23)}_{ab}  \gamma^{(3)}_{bd}} = s_2^A(a) \cX_{\gamma^{(2)}_{ac}  \gamma^{(23)}_{cd}} ~.
\eea
Thus $\alpha$ corresponds to the abelian parallel transport along the path $\gamma^{(23)}_{cd} ( \gamma^{(23)}_{ab})^{-1}$ (figure~\ref{gammax}).
This path is equivalent (for $q=1$) to the closed cycle $\gamma_x$ on $\Sigma$ used in section~\ref{secFGcoords} to define the face coordinate $x$, so we finally obtain the desired result
\bea
\alpha = \cX_{\gamma_x} = x~.
\eea

\begin{figure}[tbh]
\centering
\includegraphics[width=\textwidth]{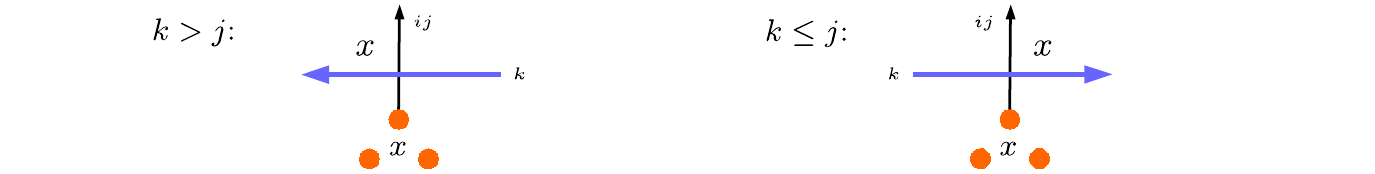}
\caption{A path on sheet $k$ crossing an $ij$-wall picks up the associated face coordinate $x$ when $k>j$ or $k\le j$, depending on whether it rotates clockwise or counter-clockwise around the puncture.}
\label{faceCoordWall}
\end{figure}

A mnemonic for the way a path picks up face coordinates as it crosses a wall is given in figure~\ref{faceCoordWall}.

%%%%%%%%%%%%%%
\paragraph{Across branch cuts:}

The transformation that reverses the orientation of a snake along an edge is given by an anti-diagonal matrix of alternating $\pm1$: 
\bea \label{Santidiag}
S = \begin{pmatrix} \cdots & 0&0&1\\ \cdots & 0&-1&0 \\ \cdots & 1&0&0  \vspace{-.17cm} \\  \iddots& \vdots & \vdots& \vdots \end{pmatrix}~.
\eea
Note that we have $S^{2} = (-1)^{K-1}$.

\begin{figure}[tbh]
\centering
\includegraphics[width=\textwidth]{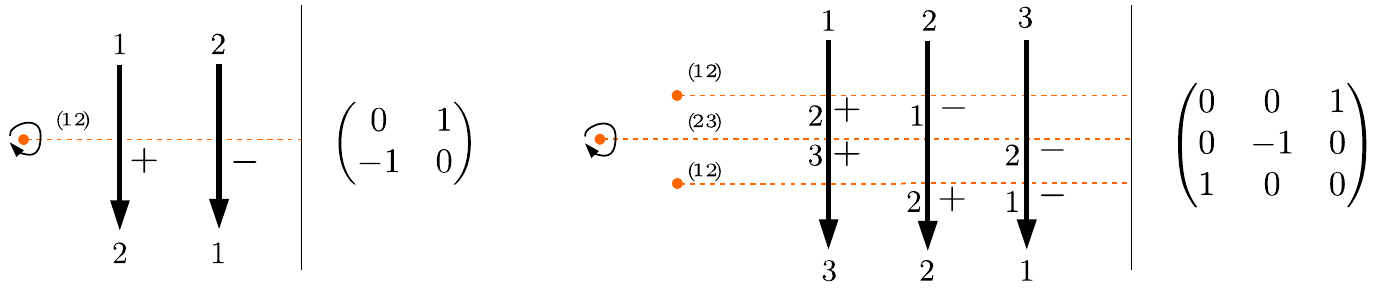}
\caption{Crossing branch cuts for $K=2$ and $K=3$ and the associated snake matrices $S$. For the reverse direction the signs are opposite.}
\label{crossingCuts}
\end{figure}

\begin{figure}[tbh]
\centering
\includegraphics[width=\textwidth]{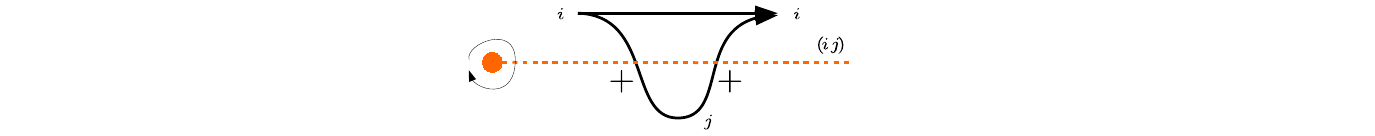}
\caption{Homotopy through a branch cut.}
\label{homotopyCut}
\end{figure}

This corresponds to crossing all the branch cuts along an edge. 
We reproduce the snake matrix~\eqref{Santidiag} by adopting the convention that crossing a branch cut of type $(i, i\pm 1)$ clockwise around its branch point produces a factor of $\pm 1$, and counter-clockwise a factor of $\mp1$ (figure~\ref{crossingCuts}).
This guarantees the invariance of parallel transport under a homotopy through a branch cut (figure~\ref{homotopyCut}).

This convention is consistent with the signs of the off-diagonal entries in the transformations~\eqref{snakeFi} and~\eqref{snakeFiInv}, since they correspond respectively to detours that circle clockwise and counter-clockwise around a branch point.
It is also consistent with the minus sign in~\eqref{XgAlmostFlat} that appears when moving through a branch point.

%%%%%%%%%%%%%%%%%%%
%%%%%%%%%%%%%%%%%%%
\section{Quantum dimension} \label{secQHol}

In the previous section we have seen how to construct the quantum parallel transport $F(\wp;q)$ along any \emph{open} path $\wp$ on $\cC$. We want to complete this construction to obtain the quantum holonomy $\Tr\, \Hol^q_\wp$ along any \emph{closed} path $\wp$. In this section we explain how this requires the introduction of a third dimension to keep track of quantum ordering. Isotopy invariance in three dimensions can then be implemented with the use of quantum group matrices.

%%%%%%%%%%%%%%
\subsection{Elevation}

Naively, we could view a closed path $\wp$ as an open path whose endpoints happen to coincide, $z_1 =z_2=z$, and compute the parallel transport as before. 
The problem is that the result would then depend on the choice of basepoint $z$. We have indeed seen in section~\ref{CoordMflat} that a left-handed crossing on $\Sigma$ contributes a factor of $q$ while a right-handed crossing contributes a factor of $q^{-1}$. The handedness is determined by considering that the segment on top is the one with the largest value of the parameter $t\in [0,1]$ along the path.
However, as we move the starting point $t=0$ through a self-intersection, its handedness changes. 
Note that even if $\wp$ itself has no self-intersection on $\cC$, a lift of $\wp$ to $\Sigma$ can have intersecting detours.

This issue can be resolved by thinking of the closed path $\wp$ on $\cC$ as a knot in the 3-manifold 
\bea
M_3=\cC\times [0,1]~. 
\eea
This idea was used by Turaev to show that the Poisson algebra generated by homotopy classes of paths on a surface $\cC$ can be quantized by the skein algebra of isotopy classes of links in the 3-manifold $M_3$~\cite{Turaev:1991}, and more recently by Bonahon and Wong to construct the $SL(2)$ quantum trace~\cite{2010arXiv1003.5250B}. 
The ordering along the interval $[0,1]$ corresponds to the quantum ordering.
We require that the quantum holonomy $\Tr\, \Hol^q_\wp$ be invariant under isotopy of $\wp$ in $M_3$, and so in particular under changes in the choice of basepoint (now understood as the lowest point of the knot along $[0,1]$).
We will find that the relative elevation along $[0,1]$ of various parts of $\wp$ is controlled by matrices of the quantum group $U_q(gl_K)$, such as the R-matrix and the cup/cap matrices.

%%%%%%%%%%%%%%
\subsection{Split triangulation}

The study of isotopy invariance for the quantum holonomy along a closed path $\wp$ in $\cC\times [0,1]$ can be reduced to a few simple configurations.
The type of isotopy transformations that we are concerned about are those that change the handedness of intersections of detoured paths in $\Sigma\times [0,1]$ (changing the handedness of a self-intersection of $\wp$ is not an isotopy). Since detours can intersect only if they are in the same triangle, we may focus on isotopy transformations taking place over a single triangle at a time.
We may further focus on changes of the relative elevation of a pair of segments of $\wp$ in the triangle, which can be iterated to produce more complicated transformations involving arbitrarily many segments.

A useful way to organize changes of relative elevation is to require that they only take place very close to the edges of the triangle. We can represent this nicely by considering a \emph{split} ideal triangulation, where the edges are thickened into ideal biangles.
The relative elevation of segments is then required to be fixed over the triangles, and change only over the biangles.
This is inspired by the procedure of Bonahon and Wong~\cite{2010arXiv1003.5250B}.

%%%%%%%%%%%%%%%%%%%%
\subsection{Isotopy invariance: R-matrix}

\begin{figure}[tbh]
\centering
\includegraphics[width=\textwidth]{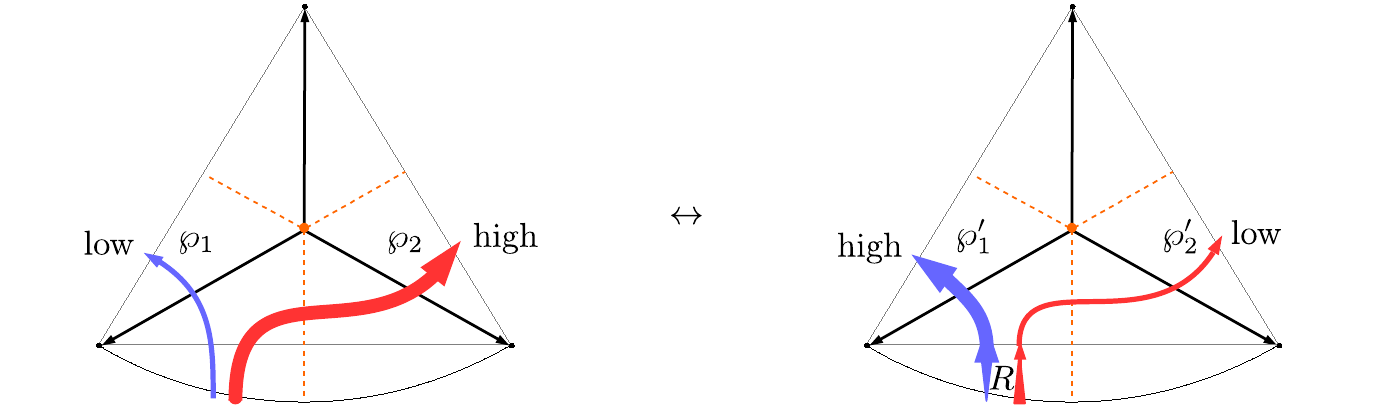}
\caption{Isotopic configurations of a pair of segments over a triangle (for $K=2$). \emph{Left:} The thin blue segment $\wp_1$ is lower along the interval $[0,1]$ than the thick red segment $\wp_2$. \emph{Right:} The relative elevation of the segments changes over the biangle, as encoded by a transformation $R$.}
\label{isotopyRmat}
\end{figure}

Let us consider two segments $\wp_1$ and $\wp_2$ that enter across the same edge of a triangle $T$ and exit across different edges (the case where they exit through the same edge can be treated similarly). 
We want to compare the case where the relative elevation of $\wp_1$ and $\wp_2$ in $T\times [0,1]$ is everywhere the same to the case where it changes. We arrange the segments such that the change of relative elevation takes place over the biangle of the split ideal triangulation that corresponds to the common edge (figure~\ref{isotopyRmat}).

\begin{figure}[tbh]
\centering
\includegraphics[width=\textwidth]{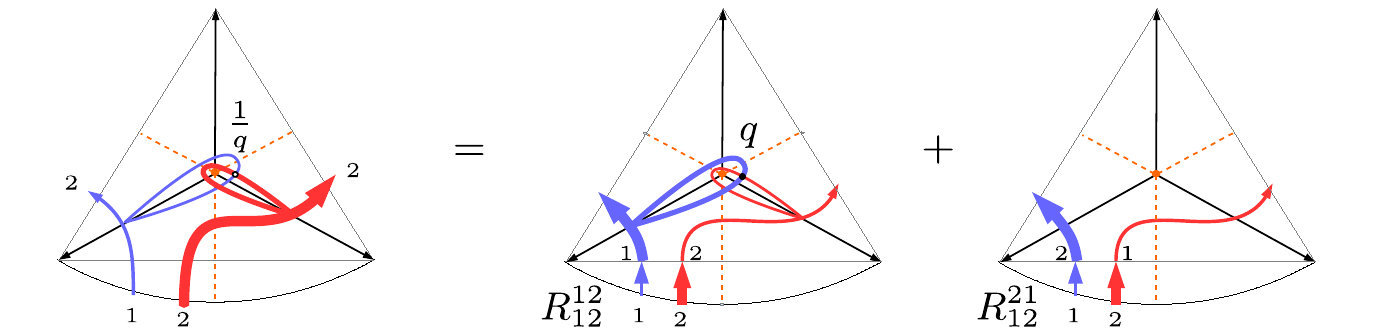}
\caption{A relation required by isotopy invariance of $\Tr\, \Hol^q_{\wp_1, \wp_2}$ involving lifts of the segments $\wp_1$ and $\wp_2$. The off-diagonal component $R_{12}^{21}$ corrects the mismatch in the handedness of the detour intersections.}
\label{isotopyR}
\end{figure}

This change of relative elevation is implemented by a transformation $R$ which acts on pairs of flat sections $s_i$ and $s_j$.
We determine $R$ by requiring isotopy invariance of the quantum holonomy $\Tr\, \Hol^q_{\wp_1, \wp_2}$.
We find that $R$ acts essentially as the identity, apart from an off-diagonal term that exchanges $s_i$ and $s_j$ for $i<j$:
\be\label{Rmatrix}
\vcenter{\hbox{\includegraphics[width=0.12\textwidth]{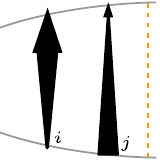}}} \qqq
R(s_i \otimes s_j) =  \left\{ 
  \begin{array}{l l }
s_i \otimes s_j  + (q^{-1} - q ) s_j \otimes s_i &   \qquad   i<j ~ ,\\
 s_i \otimes s_j  &   \qquad  i \geq j ~ ,
  \end{array} \right.
\ee
(of course, if we move the paths through the branch cut then the off-diagonal term is for $i>j$).
The role of the off-diagonal component $R_{12}^{21}$ is merely to correct the change of the handedness of the intersection of the detoured lifts of $\wp_1$ and $\wp_2$ (figure~\ref{isotopyR}). 

Correcting the elevation in the other direction (had we started with the other relative elevation on the common edge) corresponds to the inverse transformation $R^{-1}$, which simply amounts to inverting $q$:
\be\label{RmatrixInv}  
\vcenter{\hbox{\includegraphics[width=0.12\textwidth]{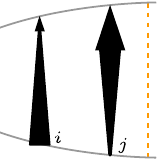}}}  \qqq
R^{-1}(s_i \otimes s_j) =  \left\{ 
  \begin{array}{l l }
 s_i \otimes s_j  + (q - q^{-1} ) s_j \otimes s_i &   \qquad   i<j~,\\
 s_i \otimes s_j  &   \qquad i \geq j ~.
  \end{array} \right. 
\ee

\begin{figure}[b]
\centering
\includegraphics[width=\textwidth]{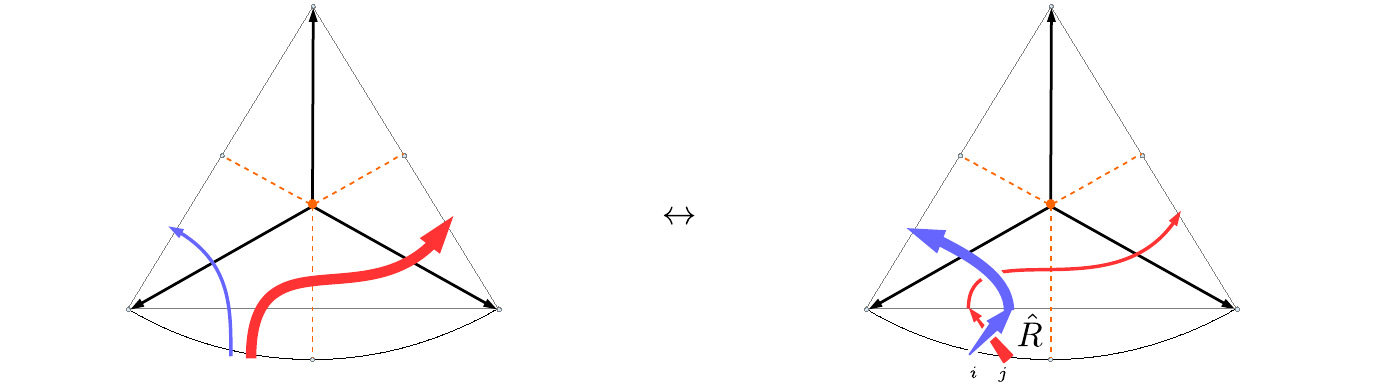}
\caption{The transformation corresponding to a crossing with a change of elevation is the $\hat R$-matrix of $U_q(gl_K)$ (here for $K=2$).}
\label{isotopyRhat}
\end{figure}

The transformation $R$ looks similar to the $\hat R$-matrix of the quantum group $U_q(gl_K)$, which acts as (we omit an overall factor of $q^{1/K}$) 
\be\label{hatRmatrix}
\hat R(s_i \otimes s_j) =   \left\{ 
  \begin{array}{l l }
 s_j \otimes s_i  + (q^{-1} - q ) s_i \otimes s_j&  \qquad i <j ~,\\
   q^{-1} s_j \otimes s_i  &  \qquad i=j~, \\
    s_j \otimes s_i   & \qquad i>j~.
  \end{array} \right. 
\ee
We can obtain a perfect match by considering a change of elevation that also crosses the segments (figure~\ref{isotopyRhat}). The segments cross again inside the triangle (without changing the relative elevation), which produces a factor of $q^{-1}$ via the abelian skein relation (figure~\ref{abelianSkein}) when the lifted segments are on the same sheet. Isotopy invariance then imposes that the transformation for a left-handed crossing with a change of relative elevation is precisely the $\hat R$-matrix:
\be
\vcenter{\hbox{\includegraphics[width=0.12\textwidth]{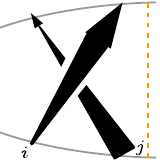}}}  \quad = \quad 
\vcenter{\hbox{\includegraphics[width=0.12\textwidth]{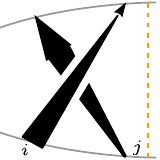}}}  \quad = \quad 
\hat R(s_i \otimes s_j)  ~.
\ee
A right-handed crossing corresponds to the inverse of the $\hat R$-matrix (obtained from $\hat R$ by inverting $q$):
\be
\vcenter{\hbox{\includegraphics[width=0.12\textwidth]{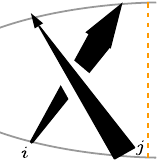}}}  \quad = \quad 
\vcenter{\hbox{\includegraphics[width=0.12\textwidth]{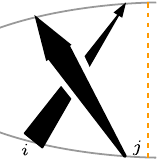}}}  \quad = \quad 
\hat R^{-1}(s_i \otimes s_j)  ~.
\ee

\begin{figure}[htb]
\centering
\includegraphics[width=\textwidth]{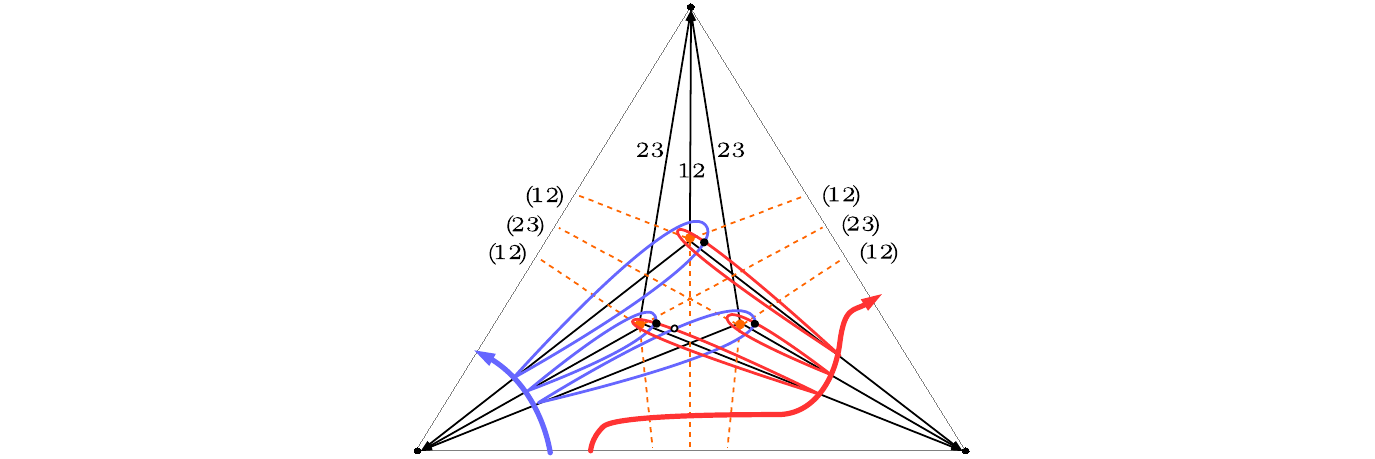}
\caption{Segments in the 3-fold cover $\Sigma$ over a triangle and their possible detours. Intersections are indicated.}
\label{specNetSL3}
\end{figure}

\begin{figure}[htb]
\centering
\includegraphics[width=\textwidth]{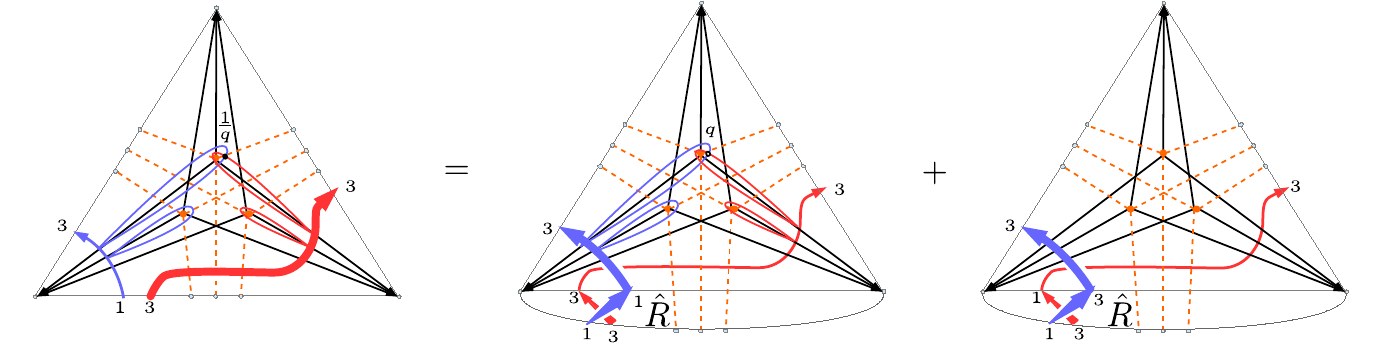}
\caption{A relation imposed by isotopy invariance of $\Tr\, \Hol^q_{\wp_1, \wp_2}$ for $K=3$.}
\label{isotopyRSL3}
\end{figure}

Although we have illustrated the calculation for $K=2$, the procedure works for any $K$. We show the general pattern of intersections of detours for $K=3$ in figure~\ref{specNetSL3}, and a sample relation imposed by isotopy invariance in figure~\ref{isotopyRSL3}.

%%%%%%%%%%%%%%%%%%%%
\subsection{Isotopy invariance: cup/cap}

\begin{figure}[tbh]
\centering
\includegraphics[width=\textwidth]{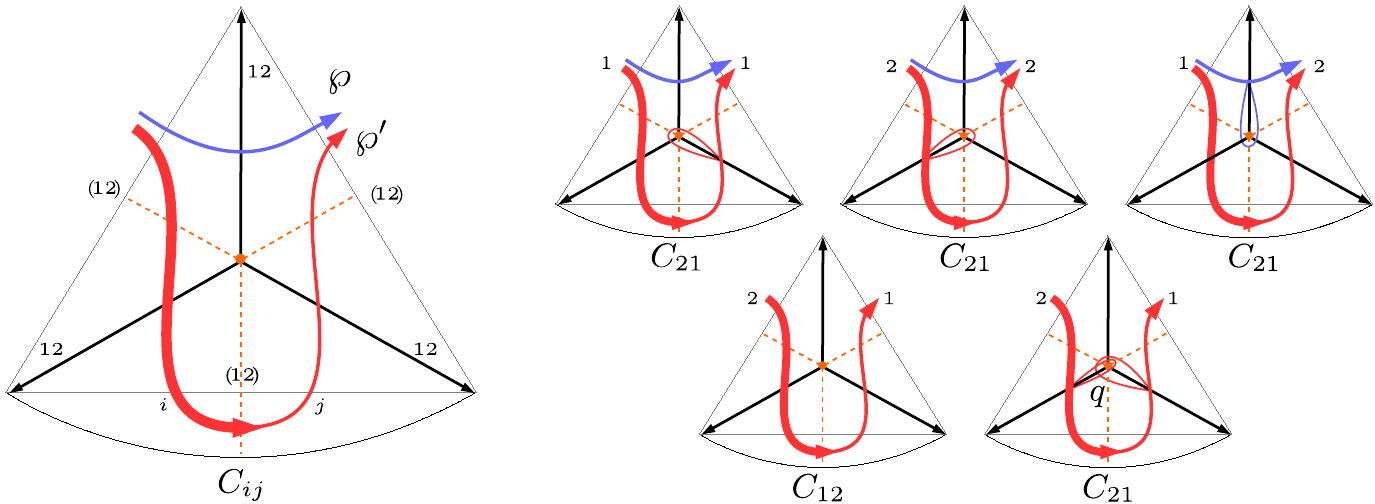}
\caption{\emph{Left}: Isotopic paths $\wp$ and $\wp'$.
\emph{Right}: The matrix elements $C_{ij}$ are determined by requiring isotopy invariance.}
\label{isotopyCup}
\end{figure}

We can also consider the case where two segments in a triangle are connected in a biangle.
We want to compare such a path $\wp$ that exits the triangle higher than it enters it, and a path $\wp'$ that exits lower than it enters. Given the interpretation of the elevation as the parameter $t\in [0,1]$ along a path, $\wp$ goes up naturally, and we can in fact use the homotopy invariance of section~\ref{secHomotopy} (as in figure~\ref{homotopyPoint}) to deform it to a path going from one edge to the next directly (figure~\ref{isotopyCup}).
In contrast, $\wp'$ requires a correction that lowers its second half compared to its first half. We therefore apply a transformation $C_{ij}$ over the biangle to correct the relative elevation of the two halves.
We can determine $C_{ij}$ by imposing that the quantum holonomy be invariant, $\Tr\, \Hol^q_\wp = \Tr\, \Hol^q_{\wp'}$. The paths shown on the top right of figure~\ref{isotopyCup} imply $C_{21}=(S^{-1})_{21}$.
The paths on the bottom right must cancel out, which gives $C_{12} = q^2 C_{21}$.

We conclude that the ``cup'' and ``cap'' transformations that lower the elevation of a path over a biangle are given by the following matrices:
\bea \label{cupCapSL2}
\vcenter{\hbox{\includegraphics[width=0.12\textwidth]{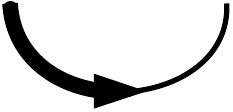}}}  \quad = \begin{pmatrix}q^{2} & 0  \\ 0 & 1\end{pmatrix} S^{-1} ~, \qqq 
\vcenter{\hbox{\includegraphics[width=0.12\textwidth]{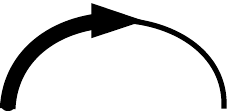}}}  \quad  = \begin{pmatrix} q^{-2} & 0  \\ 0 & 1 \end{pmatrix} S ~.
\eea
For general $K$ we find
\bea \label{cupCapSL3}
\vcenter{\hbox{\includegraphics[width=0.12\textwidth]{cup}}}  \quad &=& \text{diag} \left( q^{2(K-1)}, \ldots, q^2,1 \right)  S^{-1} ~, \nn 
\vcenter{\hbox{\includegraphics[width=0.12\textwidth]{cap}}}  \quad  &=&  \text{diag} \left( q^{-2(K-1)}, \ldots, q^{-2},1 \right)  S~.
\eea

%%%%%%%%%%%%%%%%%%%
\subsection{Skein relations and junctions} \label{secSkeinRel}

Crossings corresponding to the $\hat R$-matrices can be expressed, via the skein relation, in terms of networks with junctions (see~\cite{Tachikawa:2015iba, Coman:2015lna} for reviews).
The $\hat R$-matrix for the fundamental representation $\square$ of $U_q(gl_K)$ can indeed be decomposed as
\bea
\hat R = q^{-1} I + Q~,
\eea
where $I$ is the identity operator and $Q$ acts on a basis $s_i \otimes s_j$ of $\square \otimes \square$ as
\be\label{Qoperator}
Q(s_i \otimes s_j) =  \left\{ 
  \begin{array}{l l }
 s_j \otimes s_i -q s_i \otimes s_j  &  \qquad i <j ~,\\
0  &  \qquad i=j~, \\
    s_j \otimes s_i  -q^{-1} s_i \otimes s_j & \qquad i>j~.
  \end{array} \right. 
\ee
The operator $Q$ can be thought of as the projection $\pi_{1,1\to2}:\square \otimes \square \to \wedge^2\square$ on the second-rank antisymmetric representation, composed with the embedding $\iota_{2\to 1,1}: \wedge^2 \square \to \square \otimes \square$:
\bea 
Q =\iota_{2\to 1,1} \pi_{1,1\to2}~.
\eea
The projection and the embedding are defined by
\bea
\pi_{1,1\to2}: &&\quad  s_i \otimes s_j  \ \mapsto \  s_i \wedge s_j~, \nn
 \iota_{2\to 1,1} :  &&\quad  s_i \wedge s_j  \ \mapsto \   - q s_i \otimes s_j+  s_j\otimes s_i  \qqq i<j~,
\eea
where the $q$-deformed wedge product satisfies $s_i \wedge s_j = -q s_j \wedge s_i$ for $i<j$.
It follows that we can represent the $\hat R$-matrix graphically as the sum of a pair of parallel segments and a network with two junctions:
\bea \label{SkeinRel}
\vcenter{\hbox{\includegraphics[width=\textwidth]{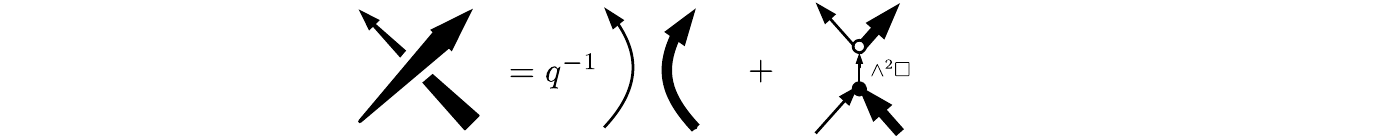}}} 
\eea
We have absorbed a factor of $q^{1/K}$ in the crossing compared to the standard form of the skein relation.

More generally, we can consider the projection from $\wedge^a \square \otimes \wedge^b \square$ to $\wedge^{a+b}\square$:
\bea
\pi_{a,b\to a+b}: \quad (s_{i_1} \wedge \cdots \wedge s_{i_a}) \otimes (s_{j_1} \wedge \cdots \wedge s_{j_b} )\ \mapsto \  s_{i_1} \wedge \cdots \wedge s_{i_a} \wedge s_{j_1} \wedge \cdots \wedge s_{j_b}~,
\eea
as well as the embedding
\bea \label{GenEmbedding}
 \iota_{a+b\to a,b} :  &&\quad  s_{k_1} \wedge\cdots \wedge  s_{k_{a+b}}  \nn 
&& \quad  \mapsto \   (- q)^{ab} \sum_{\substack{i_1 < \cdots < i_a \\  j_1 < \cdots < j_b}} (-q^{-1})^{n(i,j;k)} (s_{i_1} \wedge \cdots \wedge s_{i_a}) \otimes (s_{j_1} \wedge \cdots \wedge s_{j_b} )~,
\eea
where the sum is over disjoint splits of the indices $k_1 < \cdots < k_{a+b}$, and $n(i,j;k)$ is the minimal number of adjacent permutations to bring the sequence $i_1, \ldots, i_a,j_1, \ldots, j_b$ to $k_1, \ldots, k_{a+b}$.
This allows us to construct general networks with trivalent junctions.

An oriented path carrying some representation can be replaced with a path of reverse orientation carrying the complex conjugate representation.
Using this procedure, we can obtain junctions with only outgoing paths (source) or with only incoming paths (sink).
Classically (for $q=1$), such junctions are associated with $SL(K)$-invariant antisymmetric tensors $\epsilon_{i_1 \cdots i_K}$.
This means that each of the $K$ paths that meet at a junction lifts to a different sheet of $\Sigma$.
In the case $K=3$, we can for example replace a path labeled by $s_1 \wedge s_2$ by the reverse path labeled by $s_3$; however $s_2 \wedge s_1= -q^{-1} s_1 \wedge s_2$ should be replaced by $-q^{-1} s_3$.

%%%%%%%%%%%%%%%%%%%%
%%%%%%%%%%%%%%%%%%%%
\section{Properties of quantum holonomies} \label{secProp}

We are now ready to summarize the construction of the quantum holonomy $\Tr\, \Hol^q_\wp$.
We then discuss some of its general properties, which should be compared with the conjectured properties of the related ``quantum canonical map'' (Conjecture~12.4 in~\cite{FG} and Conjecture~4.8 in~\cite{2003math.....11245F}), as well as with the positivity conjectures for framed protected spin characters in~\cite{Gaiotto:2010be}.

%%%%%%%%%%%%%%%%%
\subsection{Quantum holonomy}

Given an oriented closed path $\wp$ on an ideal triangulation of a Riemann surface $\cC$, the quantum holonomy $\Tr\, \Hol^q_\wp$ of a flat $PGL(K,\bC)$-connection along $\wp$ can be constructed from the following steps:
\begin{itemize}
\item
choose a basepoint of $\wp$ on an edge of the triangulation,
\item
compute the quantum parallel transport $F(\wp ;q)$ along $\wp$ with the basepoint removed 
(using the associated Fock-Goncharov spectral network),
\item
close $\wp$ at its basepoint with R-matrices of the quantum group $U_q(gl_K)$.
\end{itemize}
The result does not depend on the choice of basepoint.
If $\wp$ crosses the edge with the basepoint $m$ times, the last step involves $(m-1)$ R-matrices (in particular, it is trivial if $\wp$ crosses the edge only once).
The quantum holonomy then takes the form
\bea \label{TrHolqwp2}
\Tr\, \Hol^q_\wp = \sum_\gamma\underline{\overline{\Omega}}(\wp,\gamma;q) \hat \cX_\gamma~.
\eea 
Each lift $\gamma$ of $\wp$ to the $K$-fold branched cover $\Sigma$ corresponds to a classical monomial of the form $\cX_\gamma = x_1^{a_1} \cdots x_n^{a_n}$, whose quantization is most cleanly expressed in terms of the logarithmic coordinates~\eqref{logCoordX}:
\bea \label{qtMonomial}
\hat \cX_\gamma = \exp \left( a_1 \hat X_1 + \cdots + a_n \hat X_n \right) = q^{- \sum_{\alpha <\beta}a_\alpha \varepsilon_{\alpha\beta}  a _\beta } \hat x_1^{a_1} \cdots \hat x_n^{a_n}~.
\eea
The coefficient $\underline{\overline{\Omega}}(\wp,\gamma;q) $ of this monomial is a polynomial in $q$ determined by the self-intersections of $\gamma$ and the contributions from quantum R-matrices.

%%%%%%%%%%%%%%%%%
\subsection{Left- and right-turns}

\begin{figure}[tbh]
\centering
\includegraphics[width=\textwidth]{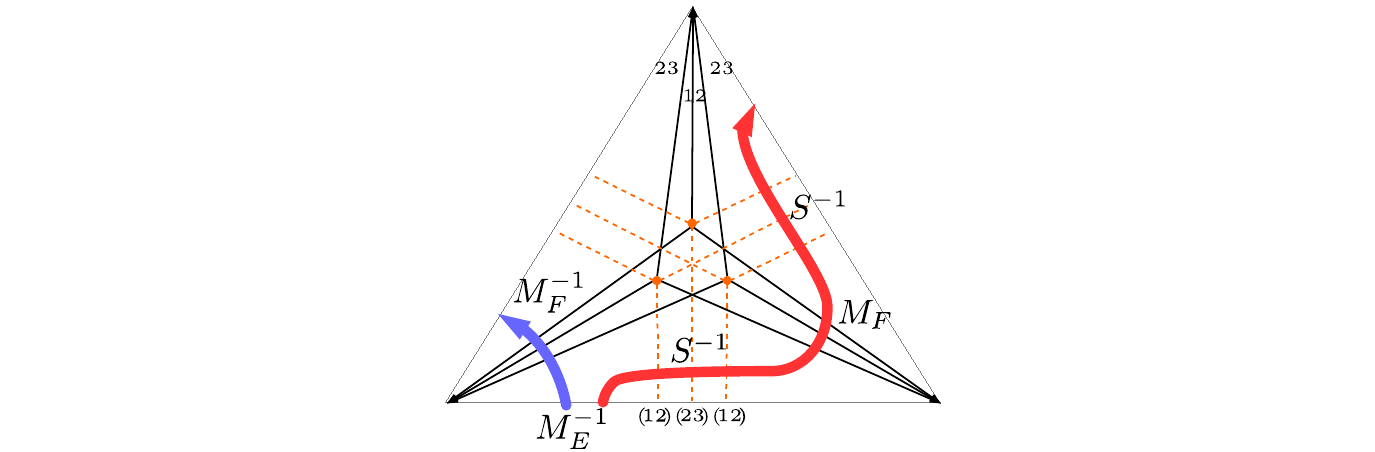}
\caption{Left- and right-turning segments in a triangle (for $K=3$). The transformation matrices across edges, cables, and branch cuts are indicated.}
\label{leftRightSegments}
\end{figure}

We can use the homotopy invariance of the quantum holonomy to put the path $\wp$ on $\cC$ in an orderly position (we assume for the moment that $\wp$ is a simple path and discuss intersections and junctions in section~\ref{secIntJunc}). We require that every segment of $\wp$ in a triangle enter across the leftmost part of an edge, and only crosses one cable of walls.
Such a segment either turns left in the triangle, in which case it does not cross any branch cut, or turns right, in which case it crosses two collections of branch cuts (figure~\ref{leftRightSegments}).
We also require that there be no cup or cap along $\wp$.

The matrix for crossing an edge with coordinates $\{ x_1, x_2, \ldots, x_{K-1} \}$ on its leftmost part is $M_E^{-1}(x_1, \ldots, x_{K-1}) $ given in~\eqref{MEinv}.
The cable transformation is $M_F^{-1}$ for a left-turn, and $M_F$ for a right-turn. The transformations across branch cuts for a right-turn are given by $S^{-1}$.
The classical holonomy is simply obtained by taking the trace of the product of snake matrices along $\wp$, which gives a Laurent polynomial in the Fock-Goncharov coordinates.

%%%%%%%%%%%%%%%%%
\subsection{Canonical lifts}

A path $\wp$ decomposed into left- and right-turns has $K$ \emph{canonical lifts} of $\wp$ to~$\Sigma$, namely the lifted paths without any detour. 
A canonical lift crosses all the edges on the same sheet.
We denote by $\wp^{(i)}$ the lift of $\wp$ that crosses all the edges on sheet $i$.
The path $\wp^{(1)}$ picks up the first eigenvalue of $M_E^{-1}$ on each edge.
It also picks up the maximal number of face coordinates, since it crosses cables counter-clockwise (around the punctures) on sheet~1, and clockwise on sheet $K$ (recall the rules in figure~\ref{faceCoordWall}). 
The canonical lift $\wp^{(1)}$ thus corresponds to the highest term, that is the monomial with the largest exponents $a_\alpha$, in the expansion~\eqref{TrHolqwp2}.
Paths $\wp^{(i)}$ with increasing values of $i$ involve fewer and fewer edge and face coordinates, up to $\wp^{(K)}$ which corresponds to the term~$\pm1$.

The coefficient of every term $\hat \cX_{\wp^{(i)}}$ in the quantum holonomy is given by  
\bea \label{coeffCanLifts}
\underline{\overline{\Omega}}(\wp,\wp^{(i)};q)  = (-1)^{n_{\text{L}}} ~,
\eea
where $n_{\text{L}}$ is the total number of left-turns along $\wp$. 
Indeed, powers of $q$ come from self-intersections or from off-diagonal components of the R-matrix.
But $\wp^{(i)}$ has no self-intersection, and it only involves the diagonal component $R_{ii}^{ii}=1$ since it always crosses edges on sheet $i$.
The sign is determined by the pattern of branch cuts that are crossed by $\wp$. Each left-turn crosses a pair of collections of branch cuts, and given that $S^{-2}=-1$, the total sign is given by the parity of $n_{\text{L}}$.

Note that it can happen that a detoured path $\gamma$ gives the same contribution as $\wp^{(i)}$, that is $\hat \cX_{\gamma} = \hat \cX_{\wp^{(i)}}$. The combined coefficient can then be a polynomial in $q$.
However, the highest term of the quantum holonomy is exactly $ (-1)^{n_{\text{L}}}  \hat \cX_{\wp^{(1)}}$. Indeed, $\wp^{(1)}$ is the unique lift of $\wp$ that receives a maximal contribution from edge and face coordinates, since any detour added to $\wp^{(i)}$ eliminates at least one face coordinate.

This should be compared to a conjecture of Fock and Goncharov, according to which the highest term $x_1^{a_1} \cdots x_n^{a_n}$ of a classical holonomy quantizes to~\eqref{qtMonomial}, with unit coefficient~\cite{FG,2003math.....11245F}.

%%%%%%%%%%%%%%%%%
\subsection{Positivity}

\begin{figure}[tbh]
\centering
\includegraphics[width=\textwidth]{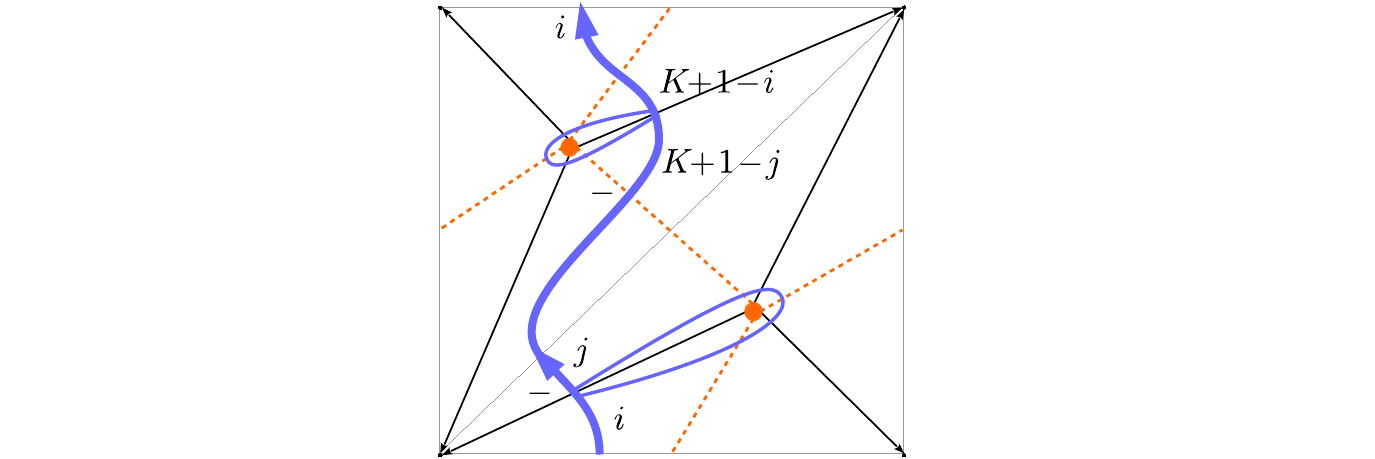}
\caption{Pair of detours $\gamma_{ij}$ and $\gamma_{K+1-j,K+1-i}$ that modifies $\wp^{(i)}$.}
\label{posDetours}
\end{figure}

All the paths $\gamma$ contributing to the quantum holonomy~\eqref{TrHolqwp2} can be obtained by gradually adding detours to the canonical lifts $\wp^{(i)}$.
Note that in order for the modified paths to close, detours must be added in pairs.
Such a pair involves a detour $\gamma_{ij}$ on a left-turn and a detour $\gamma_{K+1-j,K+1-i}$ on a right-turn (figure~\ref{posDetours}). The edges between these two detours are now crossed on sheet $j$ instead of sheet $i$, so the resulting monomial is a hybrid of the canonical lifts $\wp^{(i)}$ and $\wp^{(j)}$.
The collections of branch cuts between the two detours now contribute an opposite sign (because in our conventions $j=i+1$), and given that there is an odd number of them, the total contribution is a minus sign.
Another minus sign comes from the detour on the left-turn (recall figure~\ref{crossingInv}).
These two signs cancel each other and the overall sign of the new monomial is the same as for the original monomial.
By recursion, this implies that all monomials associated with detoured paths have the same sign $ (-1)^{n_{\text{L}}}$, as in~\eqref{coeffCanLifts}.
For paths with intersecting detours, the coefficient of each monomial can be a power of $q$.
There can be different paths leading to the same monomial, in which case their coefficients combine to produce a polynomial in $q$. 
Remark that the negative term in the off-diagonal component $(q^{-1}-q)$ of the R-matrix always cancels out against some unwanted factors of $q$ that can appear when shifting the basepoint. 

We therefore conclude that all the coefficients of the quantum holonomy are \emph{positive} Laurent polynomials in $q$:
\bea \label{positiveOmega}
\underline{\overline{\Omega}}(\wp,\gamma;q) \; \in\;  \bZ_{>0}[q,q^{-1}]
\eea
(for $n_{\text{L}}$  even, otherwise they are all negative).
This is in agreement with the ``weak positivity conjecture'' of~\cite{Gaiotto:2010be} and the conjectured positivity of the quantum canonical map in~\cite{FG}.

%%%%%%%%%%%%%%%%%
\subsection{Invariance under inversion of $q$}

\begin{figure}[tbh]
\centering
\includegraphics[width=\textwidth]{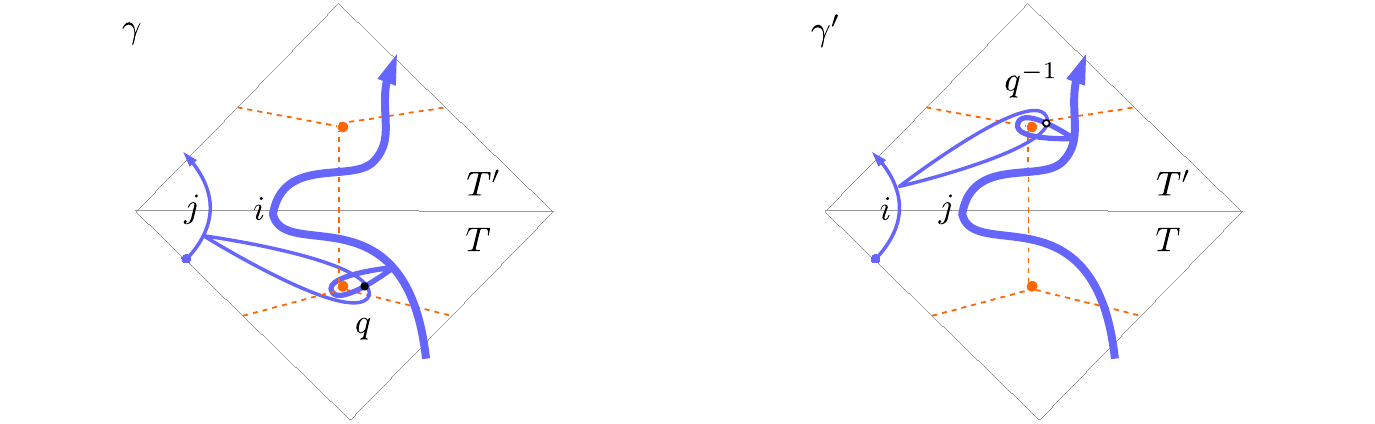}
\caption{Parts of paths $\gamma$ and $\gamma'$ with $\cX_{\gamma} = \cX_{\gamma'}$ but with different self-intersections (the paths $\gamma$ and $\gamma'$ are identical away from the two triangles $T$ and $T'$).}
\label{detourPairs}
\end{figure}

\begin{figure}[tbh]
\centering
\includegraphics[width=\textwidth]{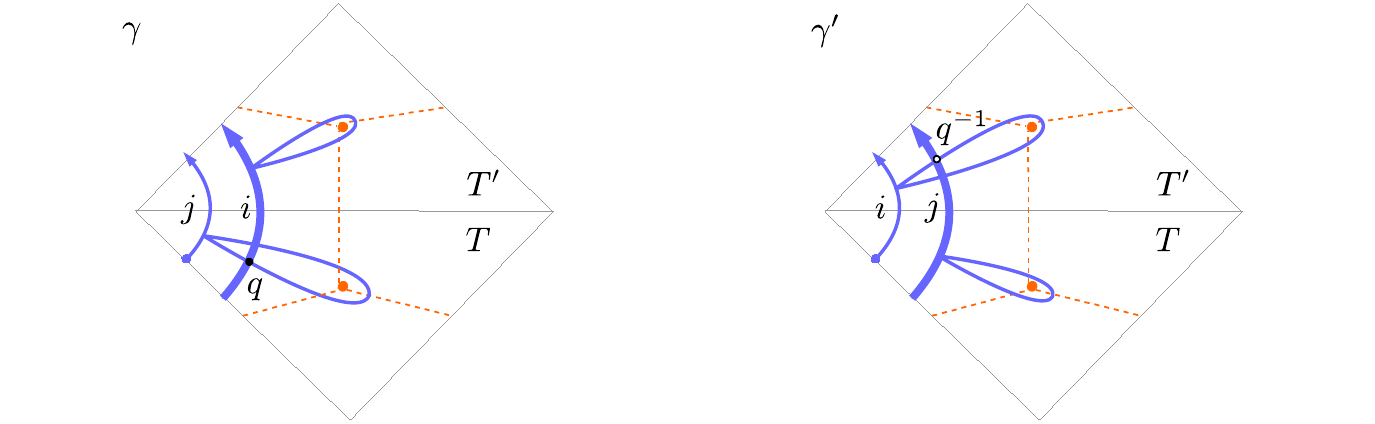}
\caption{Same phenomenon as in figure~\ref{detourPairs} but for segments turning in the same direction.}
\label{detourPairs2}
\end{figure}

Let us examine more closely the patterns in which powers of $q$ appear in the coefficients $\underline{\overline{\Omega}}(\wp,\gamma;q)$.
We assume for simplicity that there is a choice of basepoint on $\wp$ such that no R-matrix is required.
This implies that all the powers of $q$ come from intersections of detours in the lifts of $\wp$ to $\Sigma$. 
The basic contribution is a factor of $q$ coming from a right-handed intersection of two detours along two segments of a lift $\gamma$ in a triangle $T$. These two segments cross (at least) one common edge, and hence share another triangle $T'$. Then there is another lift $\gamma'$ of $\wp$ which gives the same monomial $\cX_{\gamma'} = \cX_{\gamma}$, but has a left-handed intersection of detours in $T'$ instead of a right-handed intersection in $T$ (figure~\ref{detourPairs}).
The segments of $\gamma$ cross the common edge on sheets $j$ and $i$, while the segments of $\gamma'$ cross it on sheets $i$ and $j$, so the contribution of the edge coordinates is the same in both cases. The total contribution of these two paths to the coefficient $\underline{\overline{\Omega}}(\wp,\gamma;q)$ is $q+q^{-1}$. The case where the two segments turn in the same direction is shown in figure~\ref{detourPairs2}.

More complicated paths may have many detours and contribute higher powers of $q$, but the essential phenomenon stays the same. An edge that is crossed by multiple segments of $\wp$ gives rise to a family of lifted paths with detours contributing the same monomial to the quantum holonomy. These lifted paths differ by a choice of sheet for each segment as it crosses the edge, such that there is always the same number of paths on each sheet. Each path with intersections is accompanied by a path with intersections of the reverse handedness (the detours are reflected over the edge). 
This implies that the quantum holonomy is invariant under $q\to q^{-1}$:
\bea \label{invqOmega}
\underline{\overline{\Omega}}(\wp,\gamma;q) = \underline{\overline{\Omega}}(\wp,\gamma;q^{-1})~.
\eea
This lends partial support to the ``strong positivity conjecture'' in~\cite{Gaiotto:2010be}, according to which the framed protected spin characters are linear combination of $su(2)$ characters $\chi_n(q)$ with nonnegative integral coefficients.
This is also related to the ``self-duality'' property of~\cite{FG,2003math.....11245F}.

%%%%%%%%%%%%%%%%%
\subsection{Self-intersections and junctions} \label{secIntJunc}

So far in this section we have assumed that the path $\wp$ on $\cC$ had neither self-intersection nor junction.
If $\wp$ has a self-intersection, there are additional powers of $q$ for the lifts of $\wp$ such that the two segments at the self-intersection are on the same sheet.
The coefficients~\eqref{coeffCanLifts} of the canonical lifts thus become
\bea 
\underline{\overline{\Omega}}(\wp,\wp^{(i)};q)  = (-1)^{n_{\text{L}}} q^w~,
\eea
for some integer $w$.
The positivity of the quantum holonomy is not affected by self-intersections.

However, the invariance under inversion of $q$ does not hold anymore for paths $\wp$ with self-intersections, since some powers of $q$ are shifted.
One way to restore this invariance is to apply skein relations such as~\eqref{SkeinRel} to trade paths with self-intersections for networks with junctions (and simple paths). The first term on the right-hand side of~\eqref{SkeinRel} accounts for all the lifts of $\wp$ such that the self-intersection on $\cC$ lifts to a self-intersection on some sheet of $\Sigma$. This is equivalent, up to the overall factor of $q^{-1}$, to a simple path without self-intersection.
The second term in~\eqref{SkeinRel} corresponds to lifts of $\wp$ such that the two segments at the self-intersection are on different sheets. Since these lifts do not self-intersect, the factors of $q^{-1}$ that ruin the invariance under $q \to q^{-1}$ are absent. This second term is represented by a pair of junctions.

One might worry that the various minus signs that appear in the definition of junctions, as in~\eqref{GenEmbedding}, could spoil the positivity of the quantum holonomy for a network.
Fortunately this is not the case, as we can see by rewriting the skein relation~\eqref{SkeinRel} as
\bea 
\vcenter{\hbox{\includegraphics[width=\textwidth]{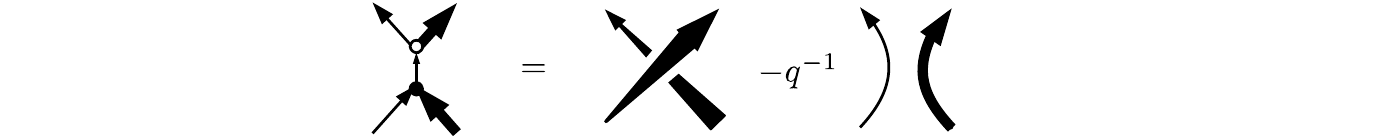}}} 
\eea
and noting that although the right-hand side is the difference of two positive polynomials, the second term is fully canceled by the first term, leaving a positive polynomial.
In summary, for networks with junctions, both the positivity property~\eqref{positiveOmega} and the invariance~\eqref{invqOmega} under inversion of $q$ hold.

%%%%%%%%%%%%%%%%%%%
%%%%%%%%%%%%%%%%%%%
\section{Examples} \label{secExamples}

We illustrate our construction by computing non-trivial framed protected spin characters associated with a line defect on the punctured torus for $K=2$, and with a pants network on a three-punctured sphere for $K=3$.

%%%%%%%%%%%%%%%%%%%
\subsection{Punctured torus}

\begin{figure}[tbh]
\centering
\includegraphics[width=\textwidth]{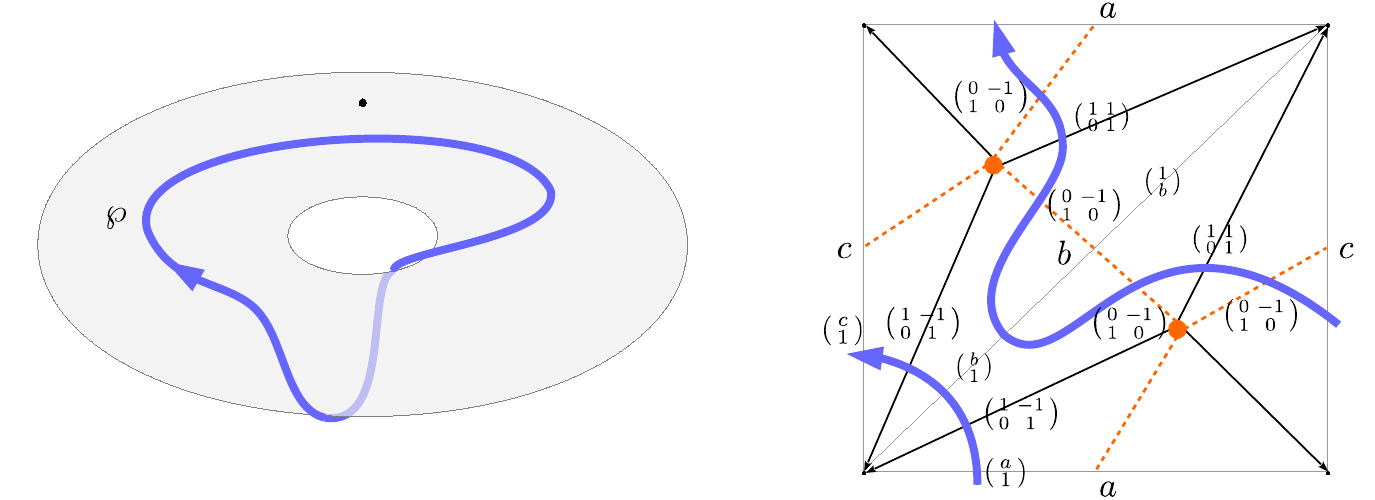}
\caption{\emph{Left:} Torus with one puncture. The path $\wp$ goes once around the A-cycle and once around the B-cycle. \emph{Right:} triangulation and spectral network for $K=2$. There are three Fock-Goncharov coordinates $a,b,c$ on the edges (the top and bottom edges are identified, as are the left and right edges). Transformations across the walls of the spectral network, the edges, and the branch cuts are indicated.}
\label{torusABloop}
\end{figure}

For our first example, we take $\cC$ to be the torus with one puncture and set $K=2$ (figure~\ref{torusABloop}).
This corresponds to the so-called $\cN=2^*$ $SU(2)$ theory, obtained from a mass deformation of $\cN=4$ super Yang-Mills theory.
The classical $PGL(2,\bC)$ holonomy along a path $\wp$ that wraps once around the A-cycle and once around the B-cycle decomposes in terms of Fock-Goncharov coordinates $a,b,c$ on the edges of the triangulation as
\bea \label{trABtorus}
\tr AB^{-1} =  1+a + 2 ab + ab^2 + ab^2 c  ~.
\eea
While the unit framed BPS indices are expected to give unit protected spin characters (because of the invariance~\eqref{invqOmega} under inversion of $q$), the coefficient of 2 should quantize in an interesting way.

\begin{figure}[tbh]
\centering
\includegraphics[width=\textwidth]{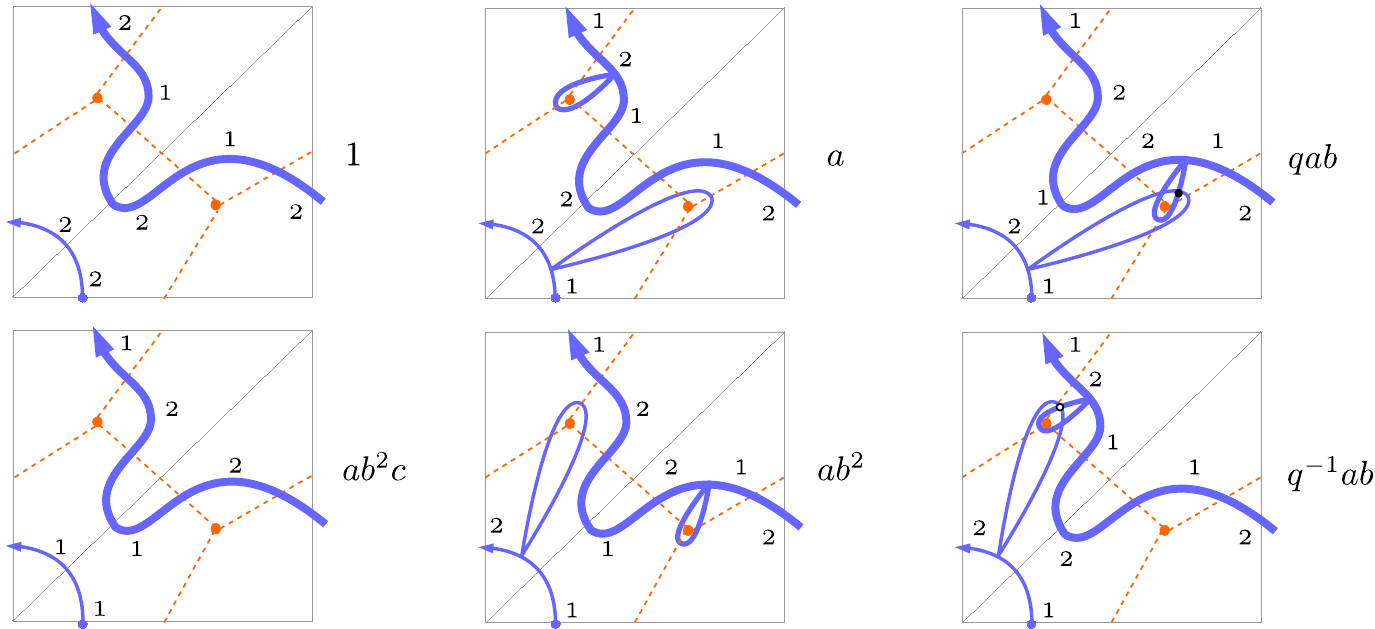}
\caption{Paths with detours contributing to the quantum holonomy along $\wp$ with basepoint on the $a$-edge (for clarity, the spectral network is omitted).}
\label{torusQuantum2}
\end{figure}

To compute the corresponding quantum holonomy, we choose a basepoint and control the elevation with $R$-matrices in order to close the path in $\cC\times [0,1]$.
Let us take the basepoint on the $a$-edge, and always move up as we go along $\wp$. 
As we come back to the $a$-edge, we can close the path by correcting the elevation with the $R$-matrix~\eqref{Rmatrix}, which in this trivial case is just the identity.
We then consider all the possible lifted paths with detours on the 2-fold cover $\Sigma$ of the punctured torus (figure~\ref{torusQuantum2}).
The paths without any detour match the lowest and highest terms $1$ and $\ex^{ \hat A + 2 \hat B + \hat C}$.
There are two paths with detours that have self-intersections, one left-handed and the other right-handed. Together they contribute $(q+q^{-1}) \ex^{ \hat A + \hat B }$ to the quantum holonomy. The framed BPS index $\underline{\overline{\Omega}}(\wp,\gamma) = 2$ is thus quantized to the framed protected spin character $\underline{\overline{\Omega}}(\wp,\gamma;q) =q+q^{-1}$. 
More explicitly, the quantum holonomy along the path $\wp$ is given by
\bea 
\Tr\, \Hol^q_\wp &=& \sum_\gamma\underline{\overline{\Omega}}(\wp,\gamma;q) \hat \cX_\gamma \nn
&=& 1 +  \ex^{ \hat A } + (q+q^{-1}) \ex^{ \hat A + \hat B } + \ex^{ \hat A + 2\hat B } + \ex^{ \hat A + 2\hat B+\hat C } ~,
\eea 
where the sum is over the paths $\gamma$ shown in figure~\ref{torusQuantum2}, with basepoint on the $a$-edge.

\begin{figure}[tbh]
\centering
\includegraphics[width=\textwidth]{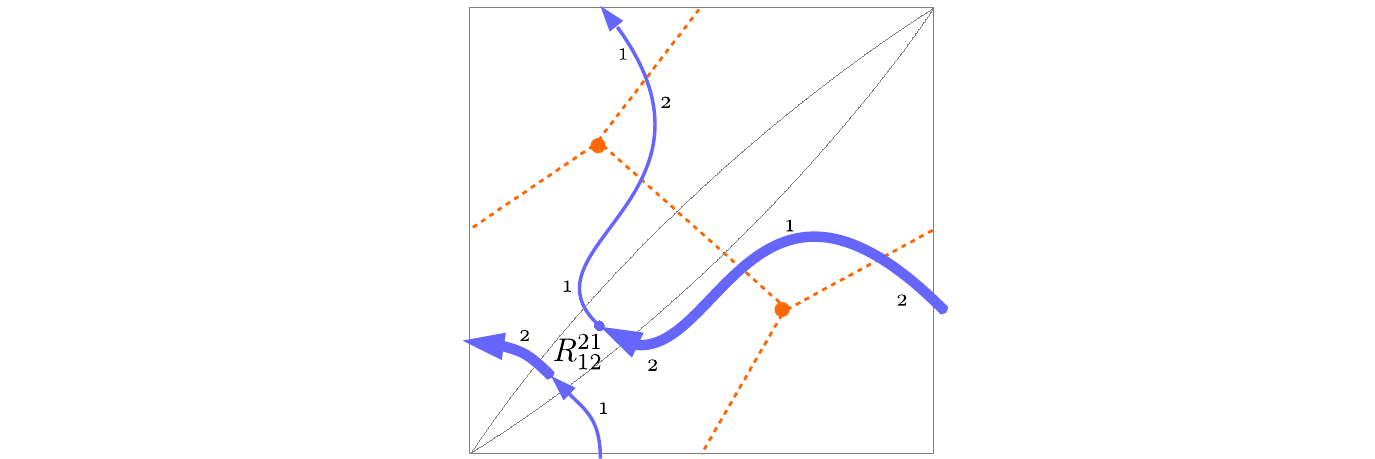}
\caption{Path contributing to the quantum holonomy when the basepoint is on the $b$-edge. The change of relative elevation needed to close the path is implemented by the quantum $R$-matrix, which has an off-diagonal component $R_{12}^{21}$.}
\label{torusQuantumRmat}
\end{figure}

How does the computation change if we decide to put the basepoint on the $b$-edge?
In order to close the path on the $b$-edge, we need to change the relative elevation of the two segments crossing it by inserting an $R$-matrix.
The off-diagonal component $R_{12}^{21}$ then allows for a new path shown in figure~\ref{torusQuantumRmat} that contributes $(q^{-1}-q)\ex^{ \hat A + \hat B }$ to the quantum holonomy.
On the other hand, the two paths with self-intersections now contribute $2q\ex^{ \hat A + \hat B }$. 
Their combination again gives the correct result.

%%%%%%%%%%%%%%%%%%%
\subsection{Three-punctured sphere}

\begin{figure}[htb]
\centering
\includegraphics[width=\textwidth]{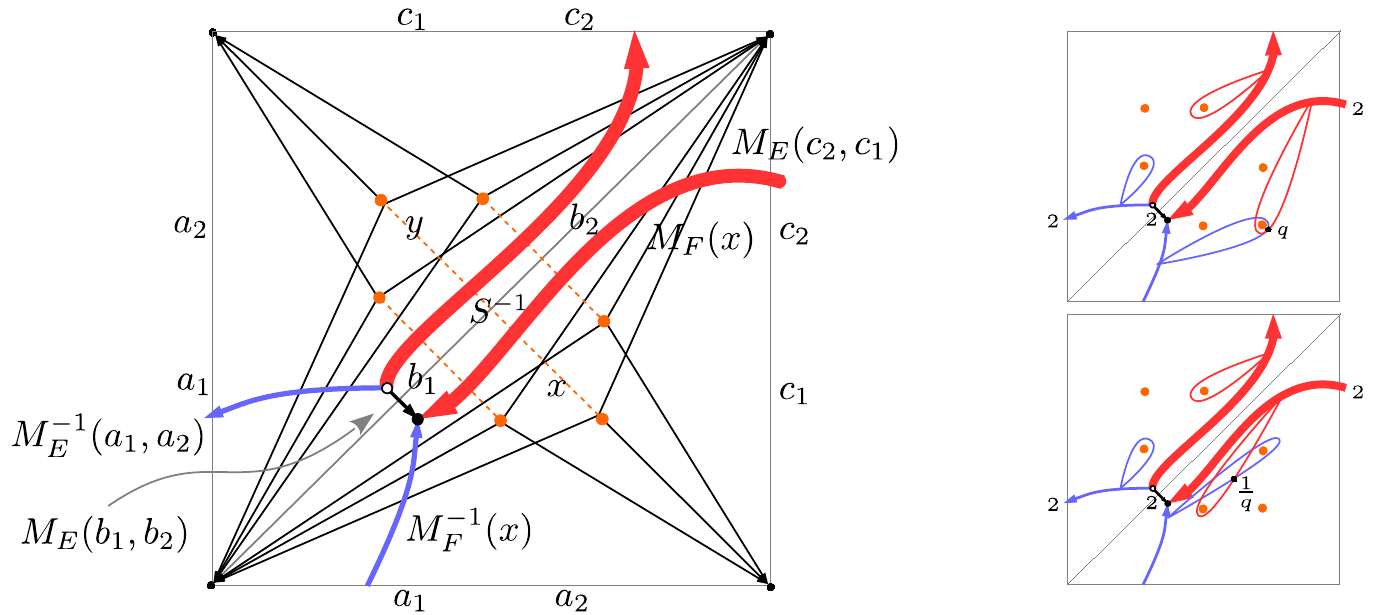}
\caption{\emph{Left}: Pants network with two junctions on the three-punctured sphere. The relevant edge and face matrices are indicated.
\emph{Right}: The two lifts of the pants network to $\Sigma$ that contribute to the framed protected spin character $\underline{\overline{\Omega}}(\wp,\gamma;q) = q+q^{-1}$.}
\label{pantsSpecNet}
\end{figure}

Another important example is the three-punctured sphere, also known as the pair of pants.
It corresponds to the theories $T_K$, which can be used as fundamental building blocks for general theories of class S~\cite{Gaiotto:2009we}.
Any simple closed curve on the three-punctured sphere is homotopic to a loop surrounding a single puncture.
However, for $K=3$ we can consider a \emph{pants network} with two junctions~\cite{Coman:2015lna}, which can be thought of as arising from the resolution via the skein relation~\eqref{SkeinRel} of a self-intersecting figure-8 loop around two punctures (figure~\ref{pantsSpecNet} left).
Using the fact that we can reverse the direction of an arrow if we simultaneously replace the representation it carries by its conjugate, we replace the small segment between the junctions with the representation $\wedge^2 \square$ by a reverse segment with the representation $\square$. 

The classical holonomy for this network has one term with a coefficient of 2.
This term corresponds to the two lifted networks with detours shown on the right of figure~\ref{pantsSpecNet}.
Given that their self-intersections have opposite handedness, we find the framed protected spin character $\underline{\overline{\Omega}}(\wp,\gamma;q) =q+q^{-1}$, in agreement with~\cite{Coman:2015lna}.

%%%%%%%%%%%%%%%%%%%%%%%%%%%%%%%%%
\section*{Acknowledgments}
It is a great pleasure to thank Dylan Allegretti, Christopher Beem, Francis Bonahon, Clay C\'ordova, Tudor Dimofte, Dmitry Galakhov, Andrew Neitzke, and Mauricio Romo for enlightening discussions.
I am very grateful to the organizers of the conference Curve 2015 at the {\it Institut de Math\'ematiques de Jussieu}, where the idea of this paper emerged. 
This work was supported by the Swiss National Science Foundation (project P300P2-158440).

%%%%%%%%%%%%%%%%%%%%%%%%%%%%%%%%%
%%%%%%%%%%%%%%%%%%%%%%%%%%%%%%%%%
%\bibliography{mybiblio}{}
%\bibliographystyle{JHEP}

\providecommand{\href}[2]{#2}\begingroup\raggedright\endgroup

\end{document}